\documentclass[]{aa}
\usepackage{hyperref}
\usepackage{comment}
\usepackage{placeins}

\usepackage{xcolor}
\hypersetup{    
    colorlinks=true,
    citecolor = blue,
    linkcolor=blue,
    filecolor=magenta,      
    urlcolor=cyan,
}
\usepackage{multirow}

\usepackage{graphicx}
\usepackage{txfonts}

 \def\mso{\,\mathrm{M}_\odot}
 \def\rso{\,{\rm R}_\odot}
 \def\lso{\,{\rm L}_\odot}

 \def\msoy{\, \mso~{\rm yr}^{-1}}
 \def\kms{\, \rm km\,s^{-1}}
 \def\gcc{\,{\rm g}\,{\rm cm}^{-3}}
 \def\Msun{\,\mathrm{M}_\odot}
 \def\Rsun{\,{\rm R}_\odot}
 \def\Lsun{\,{\rm L}_\odot}

\newcommand{\case}[1]{Case\,#1}
\newcommand{\Case}[1]{Case\,#1}

 \def\qi{q_\mathrm i}
 
 \def\days{\,\text{d}}
 \def\years{\,\text{yr}}
 \def\kyr{\,\text{kyr}}
 \newcommand{\Type}[1]{\text{Type}\,\text{#1}}
 \newcommand{\type}[1]{\text{Type}\,\text{#1}}
     \newcommand{\SN}[1]{\object{SN {#1}}}
 \newcommand{\Model}[3]{Model\,\text{#1}\text{#2}-\text{#3}}
 \newcommand{\model}[3]{Model\,\text{#1}\text{#2}-\text{#3}}
 \newcommand{\multilinecomment}[1]{}
 
\newcommand{\Fig}[1]{Fig.\,\ref{#1}}
\newcommand{\Figure}[1]{Figure\,\ref{#1}}
\newcommand{\Sect}[1]{Sect.\,\ref{#1}}

\newcommand{\Tab}[1]{Table\,\ref{#1}}

\newcommand{\App}[1]{Appendix\,\ref{#1}}
\newcommand{\Appendix}[1]{Appendix\,\ref{#1}}

\newcommand{\EDIT}[1]{#1}
\newcommand{\REFF}[1]{ }
\usepackage{soul}
\newcommand{\REM}[1]{ }
\newcommand{\EDITL}[1]{#1}

\def\simle{\mathrel{\hbox{\rlap{\hbox{\lower4pt\hbox{$\sim$}}}\hbox{$<$}}}}
\def\simgr{\mathrel{\hbox{\rlap{\hbox{\lower4pt\hbox{$\sim$}}}\hbox{$>$}}}}

\begin{document}

   \title{Interacting supernovae from wide massive binary systems}
    

   \author{A. Ercolino \inst{1}
          \and H. Jin \inst{1} 
          \and N. Langer\inst{1,2} 
          \and L. Dessart \inst{3} 
          }

   \institute{Argelander Institut für Astronomie,
              Auf dem Hügel 71, DE-53121 Bonn, Germany\\
              \email{aercolino@astro.uni-bonn.de}
         \and
          Max-Planck-Institut für Radioastronomie, Auf dem Hügel 69, DE-53121 Bonn, Germany
          \and
Institut d'Astrophysique de Paris, CNRS-Sorbonne Universit\'e, 98 bis boulevard Arago, F-75014 Paris, France
}

\date{Received August 3, 2023; Accepted January 29, 2024}
 
  \abstract
   {The features in the light curves and spectra of many \Type I and \Type II supernovae (SNe) can be understood by assuming an interaction of the SN ejecta with circumstellar matter (CSM) surrounding the progenitor star. This suggests that many massive stars may undergo various degrees of envelope stripping shortly before exploding, and may therefore produce a considerable diversity in their pre-explosion CSM properties.
   }
   {We explore a generic set of about 100 detailed massive binary evolution models in order to characterize the amount of envelope stripping and the expected CSM configurations.
    }
   {Our binary models were computed with the MESA stellar evolution code, considering an initial primary star mass of 12.6$\mso$ and secondaries with initial masses of between $\sim\,12\mso$ and $\sim\,1.3\mso$, and focus on initial orbital periods above $\sim\,500\days$. We compute these models up to the time of iron core collapse in the primary. 
    }
 {Our models exhibit varying degrees of stripping due to mass transfer, resulting in SN progenitor models ranging from fully stripped helium stars to stars that have not been stripped at all. We find that Roche lobe overflow often leads to incomplete stripping of the mass donor, resulting in a large variety of pre-SN envelope masses. In many of our models, the red supergiant (RSG) donor stars undergo core collapse during Roche lobe overflow, with mass transfer and therefore system mass-loss rates of up to $0.01\msoy$ at that time. The corresponding CSM densities are similar to those inferred for \Type IIn SNe, such as \SN{1998S}. In other cases, the mass transfer becomes unstable, leading to a common-envelope phase at such late time that the mass donor explodes before the common envelope is fully ejected or the system has merged. We argue that this may cause significant pre-SN variability, as witnessed for example in \SN{2020tlf}. Other models suggest a common-envelope ejection just centuries before core collapse, which may lead to the strongest interactions, as observed in superluminous \Type IIn SNe, such as \SN{1994W} and \SN{2006gy}.} 
{Wide massive binaries exhibit properties that \EDITL{may} not only explain the diverse envelope stripping inferred in Type Ib, IIb, IIL, and IIP SNe, but also offer a natural framework to understand a broad range of hydrogen-rich interacting SNe. On the other hand, the flash features observed in many \Type IIP SNe, such as \SN{2013fs}, may indicate that RSG \EDIT{atmospheres} are more extended than currently assumed; this  could enhance the parameter space for wide binary interaction.
}


   \maketitle
%

\section{Introduction}
 
Massive stars produce supernovae (SNe), which play a key role in regulating star formation in galaxies and the chemical evolution of the Universe.
The majority of massive stars are found in binary systems that are close enough to enable mass transfer between the component stars \citep{Sana_massive_stars_binaries}.
Mass transfer has several consequences for the SNe produced by the binary components \citep[see e.g., ][]{Podsiadlowski_massive_star_binary_interaction_1992, Wellstein_Langer, Claeys_b, Yoon_IIb_Ib, Sravan_b, Long_Binary_IIb_2022}. 

First, stellar mass is the most essential parameter in determining the fate of the star \citep[see e.g.,][]{Poelarends_stellarevo_masses_2007PhD, Smartt_rev_2009, Langer_review_2012}. Mass transfer therefore has the potential to affect the engine in the stellar core that produces the SN \citep{Brown2001_formXraybin, Laplace_core_single_vs_binary_2021, Schneider_preSNevo_stripped_2021}. Mass transfer can also drastically change the envelope properties of SN progenitors, notably their envelope mass, radius, wind, and chemical composition  \citep[e.g.,][]{Gilkis_Wind, Long_Binary_IIb_2022, Klencki_partialstripping_2022}. While these properties might not directly affect the SN engine, they determine the key observable properties of the SN \citep{1993J_Woosley, Luc2011,SN_exp_book}.

Thanks to recent large SN surveys such as PTF \citep{PTF}, Pan-STARRS \citep{PanSTARRS}, ASAS-SN \citep{ASASN}, ATLAS \citep{ATLAS}, ZTF \citep{ZTF}, and others, estimates have been derived in recent years of the properties of the envelope and circumstellar medium (CSM) of many core collapse SNe.
Notably, the progenitors of \type Ib and \Type Ic SNe have been found to be hydrogen and helium depleted, respectively.  
In both cases, the mass of the envelope above the metal core is thought to be small, and the progenitor stars are found to be rather compact.
Massive binary evolution models have been successful in reconciling the large number and properties of many \type Ib and Ic SNe \citep{Luc2011, Dessart_IIn_2015, Dessart2020_Ibc, Bersten2013_2008D, Bersten2014_iPTF13bvn, DR1, DR2}. 

Studies of the \Type IIb \SN{1993J} were amongst the first to demonstrate that binary-induced envelope stripping may be less radical in wide binaries,
as the progenitor star still contained a hydrogen mass of about $0.1\sim 0.4\mso$ in an extended red supergiant(RSG)-like envelope \citep[]{1993J_Hoflich, 1993J_Woosley, Aldering_1993J, Luc_SNe_from_extended_envelopes}. 
Progenitor evolution models for \SN{1993J} also indicated
that such stripping can occur very close in time to the explosion of the donor star \citep{1993J_Podsiadlowski, Maund, Claeys_b, Ouchi2017_IIb_RSG_progenitors, Matsuoka_Sawada_BinaryInteraction_IIP_Progenitors}.

At the same time, evidence for ejecta-CSM interaction has been inferred in numerous \Type II SNe. This interaction may be short-lived and limited to the shock breakout phase in an otherwise standard \Type IIP SN (e.g., \SN{2013fs}, \citealt{Yaron_Flash}; \SN{2020tlf}, \citealt{Jacobson_Galan_2022_precurso2020tlf}). It may also be present for a long time, producing a bona fide \Type IIn SN, either with a luminous fast-declining light curve (e.g., \SN{1998S}, \citealt{Leonard_1998S, Fassia2000_1998S, Fassia2001_1998S}) or an enduring superluminous light curve (e.g., \SN{2010jl}; \citealt{Zhang_SN2010jl, Fransson_2010jl}). Interaction may also be invisible early on but progressively \EDITL{turns on} months or years after the explosion (e.g., \SN{1993J}, \citealt{Matheson2000_1993J_I, Matheson2000_1993J_II}; \SN{2014C}, \citealt{Margutti2017_2014C}).  Despite this diversity in interaction timescales, the associated CSM is related to the mass lost from the progenitor in the final phase of evolution, which can last from weeks or months up to perhaps centuries and millennia depending on the CSM expansion rate.

Such mass-loss during the late stages of pre-SN evolution has often been described as outbursts occurring during the final evolution of single stars.
It has been suggested that these outbursts could be due to, for example, generic wave-driven envelope excitation from the vigorous neutrino-loss-dominated late burning stages \citep{Quataert_Shiode_wavedriven_winds, Fuller17_waveheating_RSG,  WuFuller21_wavedrivenoutburst}, late thermonuclear flashes in semi-degenerate cores at the lower mass end of SN progenitors  \citep{Woosley1980_TypeI_flashes, Dessart2009_1994W, Woosley_Heger_2015_SiFlash}, mergers between RSGs and neutron stars
or black holes \citep{Chevalier2012_CommonEnvelopeEvolution_SNe_Interaction},
or LBV-type eruptions in massive progenitors near the Eddington-limit \citep{SmithArnett2014_Hydroinstab_Turb_preSN}.

Here we present new MESA binary calculations for initially wide binaries, in which the large scale-height of RSG atmospheres is taken into account during the mass transfer, as described in \Sect{sec:methods}. 
We analyze these models in \Sect{sec:preSN_evolution}, and find that, in contrast to most previous calculations, any degree of envelope stripping may occur as a consequence of \Case B or \Case C Roche lobe overflow (RLOF). 
In \Sect{sec:SNE_noCSM}, we discuss the consequences of the structure of our pre-SN models, in particular their envelope mass and radius, for the
ensuing SNe, while in \Sect{sec:SupernovaCSM} we estimate the types of SNe and CSM interactions that may be expected from our models.
In \Sect{sec:SN_rates}, we derive a simple rate prediction for the
types of SNe that are predicted, and discuss their uncertainties, before concluding in \Sect{sec:conclusions}.

\section{Method and assumptions}\label{sec:methods}

We ran a series of binary star models with MESA \citep[r10398,][]{MESA_I, MESA_II, MESA_III, MESA_IV, MESA_V}, in which both stars are evolved simultaneously from the zero-age main-sequence (ZAMS) up to the core collapse of the initially more massive star (more details on the numerical settings can be found in \Appendix{sec:App_CC}). Our models are part of a larger grid of models described in Jin et al. (in prep.).

In this paper, we focus on models where the primary star (i.e., the initially more massive star) starts with $M_{1,i} = 10^{1.1}\Msun \simeq 12.6 \Msun$. This choice largely avoids the formation of semi-degenerate cores and correspondingly unstable burning episodes \citep{Woosley_Heger_2015_SiFlash, degeneracy} while the mass is still favored by the initial mass function \citep[e.g.,][]{Salpeter_IMF_55}.  
At the same time, wind stripping is negligible for stars of $12.6 \Msun$, such that corresponding single stars die with a large envelope mass, likely causing \Type IIP SNe, and any significant reduction of the envelope mass must have a binary origin. 
The primary star is in orbit with a secondary star (i.e., the initially less massive companion) with a mass ratio from  $\qi:=M_{2,i}/M_{1,i}=0.10$ to $0.95$.
The initial orbital periods chosen for our models range between $562\days$ and $2818\days$. 
Each component in the binary systems is given an initial rotational velocity of $20\%$ of its breakup velocity, corresponding to the lower velocity peak in the observed rotational velocity distribution of \cite{VLT_FLAMES_X}.

In order to identify individual evolutionary models from our set, we use a nomenclature starting with capital letters N, B, BC, and C to identify the mass transfer case of the model (N implying no interaction), followed by two numbers which identify the initial orbital period in days and the initial mass ratio, separated by a dash. For example, \model{BC}{1413}{0.80} starts with an orbital period of $1413\days$ and a mass ratio of $0.80$, and it first undergoes \Case B  and then \Case C mass transfer. In practice, a model is labeled as having undergone a phase of mass transfer if at least $0.01\Msun$ of material has been stripped from the donor star via mass transfer.

In the following, we list the most important physical assumptions of our work. A more in-depth presentation of the uncertainties is given in \Appendix{sec:APP_uncertanties}.

\subsection{Metallicity, opacities, and nuclear network}
For the initial chemical composition of our models, we adopt the values from \citet{Zsun_Asplund2021},  which estimates the proto-solar metallicity to be $Z_\odot = 0.0154$ and provides an updated table of the relative isotopic distribution of metals. The opacity tables for high temperature ($\log T \textrm{[K]} >  4$) are custom-made from the OPAL website \citep{OPAL_92, OPAL_96} and those for low temperature ($\log T \textrm{[K]} <  4$) are adopted from \cite{Ferguson2005_lowTopacity}, both sets of tables being compatible with the adopted initial chemical composition.
The nuclear network adopted is \texttt{approx21.net}, a synthetic network available in MESA and originally developed by \citet{approx19} which now also includes $^{56}\text{Fe}$ and $^{56}\text{Ni}$. The rates adopted are those from the JINA Reaclib database \citep{JINA_reaclib}. 

\subsection{Mixing and angular-momentum transport}

 Convection is implemented using the standard mixing length theory   \citep{BV_MLT, CoxBook}, with a mixing length value of $\alpha_\text{MLT}=1.5$.
 Superadiabatic layers are considered convective if the  Ledoux criterion is fulfilled, and are otherwise treated as semiconvective, with an efficiency parameter of $\alpha_\text{sc}=1$. The semiconvection parameter is motivated by the relative distribution of red and blue supergiants in the Small Magellanic Cloud \citep[SMC,][]{Abel_semiconvection}.
 Mass-dependant convective overshooting of the hydrogen-burning core is included as in \cite{Hastings_BeFractions}, extending the convective cores of our $12.6\mso$ primaries by $\sim 0.18$ pressure scale heights.
 Thermohaline mixing is included as in \citet{Cantiello_Thermohaline} with an efficiency parameter of $\alpha_\text{th}=1$.
Rotationally induced mixing and angular momentum transport are treated as in \citet{Brott2011} and \cite{fmu_Yoon2006}, and tidal torques are included as in \citet{Detmers_Tides}.
 
\subsection{Stellar wind mass-loss}
\label{sec:wind_method}
The wind prescription used is a combination of different wind recipes\EDIT{, as described in \citet{Pauli_LMC}, which is a re-elaboration of the recipe from \cite{Brott2011}. Below, we summarize the main features}. 

We define a `cold' wind for surface $T_\text{eff}$ below the bi-stability jump temperature \citep[$\simeq 25\ 700 \text{K}$, from][]{Vink2001}, in which case the maximum between the result from the prescriptions of \citet{Nieuw_wind} and \citet{Vink2001} is used. We define a `hot' wind for higher temperatures than the bi-stability jump which uses different recipes depending on surface hydrogen mass fraction $X_\text s$. If $X_\text s\geq0.7$, then the recipe from \citet{Vink2001} is used. 
For $0.4\geq X_\text s\geq0.2$, \EDIT{the mass-loss rate from \citet{NugisLamers2000} enhanced by a factor of about 2 is used \citep[this is to consider the clumping factor of about 4; see][]{Yoon_wind}} and for $X_\text s\leq 10^{-5}$ the recipe from \citet{Yoon_wind} is used, which is itself a compilation of the recipes of \cite{TSK_wind}, \cite{Hamann2006_Potsdam} and \cite{Hainich_WRwinds}. For $X_\text s$ between the specified ranges, a linear interpolation between the two closest `hot' recipes is performed. If $T_\text{eff}$ is found within $\pm5\%$ of the bi-stability jump temperature, the `cold' and `hot' recipes are linearly interpolated as a function of temperature. 

Wind mass-loss is also enhanced due to rotation via the term $\left(1- v/ v_\text{crit}\right)^{\EDIT{-0.43}}$ where $v$ is its equatorial rotational velocity and $v_\text{crit}$ is its breakup velocity \citep{rot_enhanced_wind}.

\subsection{Roche lobe overflow}
\label{sec:RLOF_method}
The choice of the initial orbital periods of our binary models is such that the primary star fills its Roche lobe climbing the RSG branch. This may happen following core-hydrogen exhaustion leading to \case B mass transfer (or Roche lobe overflow, RLOF) and after core-helium exhaustion leading to \case C. Depending on the initial conditions, our models exhibit either \case B, \case C, or both
mass transfer events.

The prescription we used to model mass transfer is that of \citet{Kolb_scheme}, which we refer to below as the Kolb scheme. This scheme was developed to deal with mass transfer from stars with extended atmospheres.
It differs from the schemes adopted in many previous binary model calculations; for example, \cite{Claeys_b} \citep[which adopts the mass transfer scheme from][]{deMink07}, \cite{PABLO_LMC_GRID}, and \cite{CHEN_SMC_GRID_MS_morph}. The latter two use MESA's default mass transfer scheme \citep{MESA_III} for systems similar to those investigated in this paper where mass is removed from the donor star until it underfills its Roche lobe. When this scheme is used for RLOF from stars with a deep convective envelope, which expand upon mass-loss \citep{rad_response_massloss_conv}, it often encounters numerical difficulties. 

The Kolb scheme does not exhibit these problems as the algorithm evaluates mass removal in a different way. Firstly, it assumes an optically thin and isothermally expanding flow \citep[as in][]{Ritter_scheme} above the photosphere, which allows mass transfer to smoothly activate before the donor stars actually begin overflowing their Roche lobes. Secondly, this scheme does not prevent the star's radius to exceed its Roche lobe radius, thus removing the rigid condition set in by the default scheme. This actual overflow of the Roche lobe occurs due to the finite speed of mass transfer. \cite{Pavlovskii_Ivanova_2015_MT_from_Giants} found a similar behavior in MESA models with their own mass transfer prescription.

We compute the angular momentum gain by the accreting star depending on whether the material is accreted via a Keplerian disk or by direct impact \citep{j_accretion}. \EDIT{The mass transfer efficiency, \EDITL{that is} the ratio of the accreted mass to the mass transferred by the donor, is not a free parameter in our models, but determined by the spin of the accretor.  Accretion is assumed to be efficient until the mass gainer approaches critical rotation, after which the mass transfer becomes inefficient  \citep{Petrovic2005_WRO_RLOF_constraint}.}

\EDIT{As discussed by \citet{Langer_review_2012}, the assumption of the donor spin regulating the mass transfer efficiency is uncertain. In particular for disk accretion, \citet{PophamNarayan1991_Accretion_stops_at_breakup} and \citet{Paczynski1991_polytrpic_accretion_disk} argue that an accretion disk may drain any surplus angular momentum form the accretion star and move it outward. \citet{Vinciguerra_2020_BeXray_Binaries} argue that indeed the properties of the Galactic Be/X-ray binary population can be best understood if the Be stars in these binaries accreted with an efficiency of at least 30\%.}

\EDIT{Our paper is concerned with the widest interacting binaries, which are not expected to form Be/X-ray binaries. Given that many of the widest binaries are likely eccentric at the time of interaction (cf.\,Sect.\,\ref{sec:initial_eccentricity}), the accretion flow may be complex and non-stationary. \cite{Walder2008_RSOph_Nova_accretion} and \cite{BoothMohamedPods_2016_innertorque_CSM_RSoph_Ia} find in 3D-SPH model of accretion from a low mass red giant to a white dwarf in a circular binary that the mass transfer efficiency is less than 10\%, which supports the
low efficiencies obtained in our models.} 

\EDIT{The fraction of the transferred mass which is not accreted by the mass gainer is assumed to leave the binary system. In our models, we assume that it carries the specific orbital angular momentum of the mass gainer.  This assumption is also uncertain, and may affect the evolution of the binary separation \citep{Podsiadlowski_massive_star_binary_interaction_1992}.
We come back to this point in Sect.\,\ref{sec:interaction_with_RLOF_CSM},
and provide a more in-depth discussion of the effects of our different assumptions in Appendix \ref{sec:APP_mass_transfer_efficiency} and \ref{sec:APP_angular_momentum_loss}. }

\subsection{Unstable mass transfer}\label{sec:unstable_RLOF}

Our models feature a donor star with an extended and deep convective envelope by the time they undergo RLOF, and as such they expand as mass is removed. 
For donors which are more massive than their companions, which is the typical situation at first, the orbit shrinks upon mass transfer.
This can lead to a runaway mass transfer, \EDITL{that is} unstable mass transfer, as more of the donor star will be overflowing the shrinking Roche lobe, where the donor's envelope will engulf the secondary star, thus starting a common-envelope evolution.

\subsubsection*{Instability criterion}

We determine whether a model system is undergoing unstable mass transfer by pinpointing whether the primary star overflows its outer Lagrangian point, which is roughly at $1.3R_\text L$ \citep{Pavlovskii_Ivanova_2015_MT_from_Giants, Pablo_Kolb}. Such outflow, which is not accounted for in the scheme developed by \citet{Kolb_scheme}, not only removes additional mass but also carries away large amounts of angular momentum, given its bigger distance from the center of mass.
How much material needs to be removed in this way to trigger a common envelope phase is not known. We define a system to undergo unstable mass transfer if more than $1\Msun$ is lost from the donor star while it exceeds its outer Lagrangian point. \EDIT{This method only reflects that 1D modeling is breaking down \citep{Temmink_2023_stability_of_MT}, rather than truely capturing the onset of unstable mass transfer. We discuss the effects of the implementation of different critaria for mass transfer stability in \App{sec:APP_RLOFstability}}.

\subsubsection*{Common envelope evolution}

MESA is a 1D code that is not well suited to describing the engulfment of a main-sequence star into a red supergiant envelope.  
For our models, we use simple prescriptions to describe the outcome and timescales of the common envelope phase \citep[see][ for more details]{Ivanova_book}.
We implement the energy criterion \citep[or the $\alpha\lambda$ criterion, ][]{Webbing_alphalambda, deKool_CEE, DewiTauris_CEE}, which assumes that a fraction $\alpha$ of the released orbital energy $\Delta E_\text{orb}$ is used to expand the envelope, which has an initial binding energy of  
\begin{equation}
    E_\text{CE} = \frac1\lambda\frac{GM_\text{CE}M_1}{R_\text{CE}},
\end{equation}
\EDIT{where $\lambda$ is a numerical factor that accounts for the density distribution of the envelope, as well as for the thermal and recombination energies \citep{Han_Lambda_Eint}.
 We compute the value of $\lambda$ following \cite{Wang_CEE} at the moment of the onset of unstable mass transfer (i.e., $M_\text{CE}=M_\text{1,env}$ and $R_\text{CE}=R_1$).} 
 The efficiency parameter $\alpha$ is poorly constrained, and we adopt $\alpha=1$, as often assumed. The condition for common envelope ejection is then
\begin{equation}
\alpha\left(\frac{GM_\text{1,core}M_2}{2a_\text{post}}-\frac{GM_1M_2}{2a_\text{pre}} \right) > \frac{GM_1M_\text{1,env}}{\lambda R_1}.
\end{equation}
It implies that the orbital separation below which enough energy will be supplied to the common envelope to unbind it from the system is
\begin{equation}
a_\text{post} =    a_\text{pre}\frac{M_\text{1,core}}{M_\text{1}}\left(\frac{2}{\alpha\lambda} \frac{M_\text{1,env}}{M_2}\frac{a_\text{pre}}{R_1}+1 \right)^{-1}.
\end{equation}

The in-spiraling system may merge before reaching this orbital separation, as the main-sequence star may fill its Roche lobe and undergo another phase of unstable RLOF, causing the two stars to coalesce. Assuming that the primary's helium core and the secondary star remain at a constant radius during this process, we can evaluate whether they will fill their new Roche lobes once the orbit size shrinks to $a_\text{post}$. We define mergers as those systems undergoing common envelope phase in which the main-sequence star fills its new Roche lobe once the system has reached the new orbital separation $a_\text{post}$. If that is not the case, we assume that the system has enough energy to eject the CE.

\EDIT{The common envelope phase of a given binary system may
    consist of a sequence of several different events, each of which evolves on its own timescale \citep{Ivanova2013_CEE_REV}. There is an initial inspiral phase in which the
companion enters the low density parts of the envelope of
the red supergiant. \citet{Betelgeuse_merger} derive
the timescale of this phase, for binaries that are similar to those discussed here, as the ratio of the orbital angular momentum to the torque on the binary, which yields
\begin{equation}\label{eq:tCE}
\tau_\text{CE} \sim \frac{8}{3c_\text D}\left(\frac{q}{1+q}\right)^\frac 32 \left(\frac{R_1}{R_2}\right)^\frac 7 2\left(\frac{R_2^3}{GM_2}\right)^\frac 1 2.\end{equation}
When all the values are taken at the moment of the onset of common envelope evolution, and the drag coefficient $c_\text D$ is set to $1$ as recommended, for a $M_1=12\Msun$ RSG with a radius of $800\Rsun$ capturing a $M_2=6\Msun$ main-sequence star with $5\Rsun$, the predicted in-spiral time is $\sim 6\,000\,$yr.}

\EDIT{Subsequently, during the so called plunge-in phase, the bulk of the energy is injected into the common envelope on a shorter timescale, short enough to be modeled by 
hydrodynamic calculations. If an envelope ejection is occurring, much of it may happen during this phase.} 
\citet{Lau_Hydro_12M}, for a model system similar to ours, find a plunge-in phase that lasts for $\sim 30\text{yr}$ and brings the cores to an orbital separation of roughly $40\Rsun$. 
In their hydrodynamic models, a major fraction of the envelope is found unbound after the plunge-in,
i.e., $\sim 95\%$ in a lower, $\sim 75\%$ in a higher resolution calculation.
However, significant fractions of the envelope may fall back later on, and the ejection
of the rest of the envelope could take place in several steps on a much longer timescale \EDIT{\citep[e.g., ][]{Passy2012_CEE_RG_SPH, Clayton2017_EpisodicME_from_CE}}. 

\EDIT{For cases where the majority of the envelope is ejected after plunge-in, still significant orbital shrinkage and mass ejection is expected. In particular, the radiative part of the RSG envelope may undergo expansion and strong interaction with the companion \citep[e.g.,][]{Pablo_Kolb, Hirai_CE_2step, Gagnier_post_plunge_in}, occurring on the thermal timescale of the remaining RSG envelope, which is typically of several hundred years.}

\EDIT{The overall picture is too complex to yield reliable predictions for those of our binary systems for which a common envelope evolution is expected. Below, we provide a comparison of the remaining lifetime of the RSG component of our models with the spiral-in time given by  Eq.\,\ref{eq:tCE}. }

\section{Presupernova evolution} \label{sec:preSN_evolution}

\begin{figure*}
       \resizebox{\hsize}{!}{\includegraphics[]{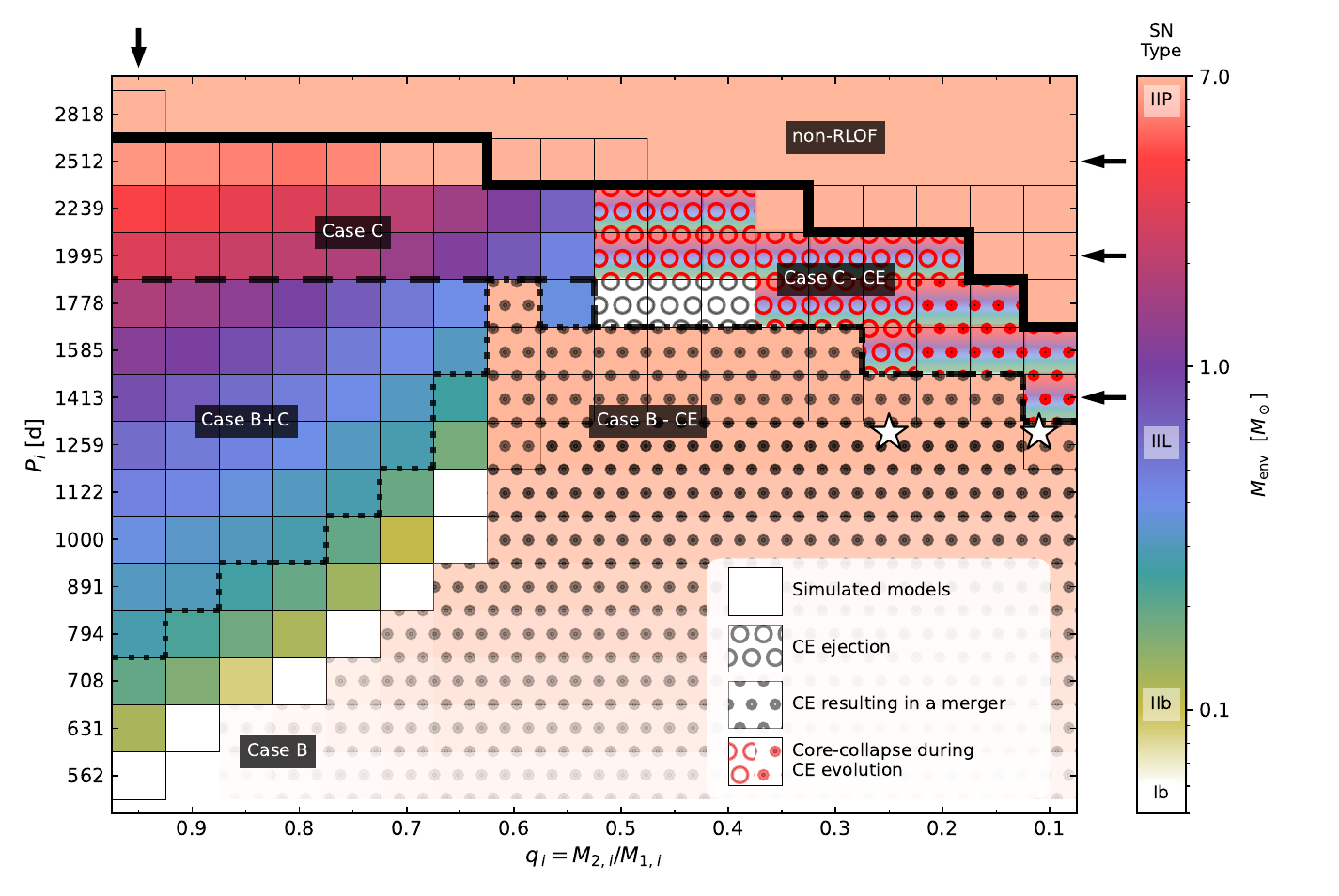}}
\caption{Initial orbital period $P_\text i$ versus the initial mass ratio $\qi$ for the computed binary evolution models (squares), with a color-coding denoting the remaining envelope mass of the primary star at the time of its core collapse. The envelope mass can be used to indicate the anticipated SN type if no CSM interaction would occur (see color bar). 
The black lines separate systems without mass transfer (above the thick full-drawn), \Case C (above the dashed line), \case B+C systems (above the dotted line), and pure \case B systems. 
where the latter is not investigated in the area in which no binary models are computed. The background pattern of
circles indicates a likely CE evolution,
where for wide, open circles the energy criterion implies 
that a CE ejection would be possible, and small filled circles 
imply the opposite. Here, symbols in red color indicate that the donor star's core may collapse before the end of the common envelope phase is reached. As in this case, the pre-SN envelope structure is not well constrained, we give the corresponding squares multiple background colors, indicating the range of possibilities.
 The star symbols represent the multi-D models of \citet{Lau_Hydro_12M}. Black arrows point out rows and columns in the plot for which data is displayed in Table \ref{table:data}. 
 }
    \label{fig:Pq-plot}
\end{figure*}

\begin{table*}
\caption{Properties of selected models \EDITL{which are highlighted in Fig.\,\ref{fig:Pq-plot}}.
}             
\label{table:data}      
\centering          
\resizebox{\textwidth}{!}{
\begin{tabular}{cccccccccccccccccc}
\hline
$P_i$ & $\qi$ & RLOF & $L$ & $ T_\text{eff}$  &  $R$    & $\mathcal{L}$ & $Y_\text s$ 
 &  $M_\text{He-core}$ & $M_\text{env}$   & $M_\text{H}$  & $\Delta M_\text{RLOF,C}$ & $\left.\frac{\dot M_\text{RLOF,C}}{\dot M_\text{wind}}\right|_\text{CC}$ & $\dot M_\text{RLOF,C}^\text{max}$ & $\Delta t_\text{RLOF,C}^\text{max}$ & $\tau_\text{CE}$ &  SN & ${E_\text{interact}}/{E_\text{LC}}$ \\
d &  &   & k$\lso$ & kK  &  $\Rsun$    & k$\mathcal{L}_\odot$ &   
 &  $\mso$ & $\mso$   & $\mso$  & $\mso$ &   & $\mso/\text{yr}$ & kyr  & kyr &   Type & \\
\hline
(1) & (2) & (3) & (4) & (5) & (6) & (7) & (8) & (9) &(10) & (11) & (12) & (13) & (14) & (15) & (16) & (17) & (18) \\
\hline
$   562\tablefootmark{*} $ & $0.95$ &    B & $ 49.2 $ & $ 24.7 $ & $   12.1 $ & $ 14.4 $ & $  0.98 $ & $  3.42 $ & $  0.00 $ & $  0.00 $ & $  0.00 $ & n.a. & n.a. & n.a. & n.a. & Ib & n.a. \\
$   631 $ & $0.95$ &      B & $ 60.5 $ & $  4.45 $ & $  415 $ & $ 15.9 $ & $  0.71 $ & $  3.67 $ & $  0.12 $ & $  0.02 $ & $  0.00 $ & n.a. & n.a. & n.a.  & n.a. &IIb & n.a. \\
$   708 $ & $0.95$ &      B & $ 64.0 $ & $  3.69 $ & $  620 $ & $ 16.3 $ & $  0.48 $ & $  3.73 $ & $  0.19 $ & $  0.06 $ & $  0.00 $ & n.a. & n.a. & n.a. & n.a.  & IIb & n.a.\\
$   794 $ & $0.95$ &     BC & $ 65.1 $ & $  3.46 $ & $  710 $ & $ 16.0 $ & $  0.44 $ & $  3.78 $ & $  0.28 $ & $  0.12 $ & $  0.01 $ & $  0.44 $ & $ 2.1\times 10^{-6} $ & $  2.73 $ & n.a.  &IIb & 0.02 \\
$   891\tablefootmark{*} $ & $0.95$ &     BC & $ 65.1 $ & $  3.39 $ & $  741 $ & $ 15.7 $ & $  0.41 $ & $  3.82 $ & $  0.31 $ & $  0.15 $ & $  0.06 $ & $  1.03 $ & $ 8.1\times 10^{-6} $ & $  4.31 $ & n.a. &IIL & 0.46 \\
$  1000 $ & $0.95$ &     BC & $ 62.1 $ & $  3.30 $ & $  763 $ & $ 14.6 $ & $  0.39 $ & $  3.86 $ & $  0.39 $ & $  0.20 $ & $  0.17 $ & $  1.70 $ & $1.2\times 10^{-5} $ & $  3.25 $ & n.a. &IIL & 1.7 \\
$  1122 $ & $0.95$ &     BC & $ 67.0 $ & $  3.27 $ & $  806 $ & $ 15.4 $ & $  0.37 $ & $  3.88 $ & $  0.46 $ & $  0.25 $ & $  0.26 $ & $  2.45 $ & $1.5\times 10^{-5} $ & $ 15.2 $ & n.a.  &IIL & 2.5 \\
$  1259 $ & $0.95$ &     BC & $ 67.8 $ & $  3.22 $ & $  839 $ & $ 15.1 $ & $  0.36 $ & $  3.88 $ & $  0.60 $ & $  0.34 $ & $  0.39 $ & $  2.77 $ & $ 2.1\times 10^{-5} $ & $ 18.4 $ & n.a.  & IIL & 3.4\\
$  1413 $ & $0.95$ &     BC & $ 68.2 $ & $  3.15 $ & $  878 $ & $ 14.4 $ & $  0.34 $ & $  3.90 $ & $  0.84 $ & $  0.50 $ & $  0.66 $ & $  3.58 $ & $ 3.5\times 10^{-5} $ & $ 21.8 $ & n.a.  & IIL & 5.2\\
$  1585 $ & $0.95$ &     BC & $ 69.2 $ & $  3.09 $ & $  917 $ & $ 13.5 $ & $  0.32 $ & $  3.92 $ & $  1.18 $ & $  0.73 $ & $  1.06 $ & $  5.18 $ & $ 5.9\times 10^{-5} $ & $ 23.7 $ & n.a.  & IIP & 7.1\\
$  1778 $ & $0.95$ &     BC & $ 70.0 $ & $  3.05 $ & $  950 $ & $ 12.3 $ & $  0.31 $ & $  3.95 $ & $  1.73 $ & $  1.11 $ & $  1.90 $ & $  6.83 $ & $ 1.2\times 10^{-4} $ & $ 25.8 $ & n.a.  & IIP &  9.9\\
$  1995 $ & $0.95$ &      C & $ 71.5 $ & $  3.03 $ & $  970 $ & $ 10.5 $ & $  0.30 $ & $  3.98 $ & $  2.80 $ & $  1.85 $ & $  4.06 $ & $ 13.0 $ & $ 5.9\times 10^{-4} $ & $ 21.2 $ & n.a.  &IIP &  14\\
$  2239 $ & $0.95$ &      C & $ 71.1 $ & $  3.05 $ & $  958 $ & $  9.18 $ & $  0.30 $ & $  3.98 $ & $  3.74 $ & $  2.50 $ & $  3.11 $ & $ 20.1 $ & $ 4.2\times 10^{-4} $ & $ 11.5 $ & n.a. & IIP &  10\\
$  2512\tablefootmark{*} $ & $0.95$ &      C & $ 71.1 $ & $  3.12 $ & $  916 $ & $  7.12 $ & $  0.30 $ & $  3.98 $ & $  5.97 $ & $  4.03 $ & $  0.88 $ & $ 80.4 $ & $ 5.5\times 10^{-4} $ & $  1.12 $ & n.a.  & IIP &  2.6\\
$  2818 $ & $0.95$ &     No & $ 71.1 $ & $  3.15 $ & $  898 $ & $  6.54 $ & $  0.30 $ & $  3.98 $ & $  6.86 $ & $  4.64 $ & $  0.00 $ & n.a. & n.a. & n.a. & n.a.  & IIP & n.a.\\ 
\hline 
$  2512 $ & $  0.90 $ &      C & $   70.9 $ & $   3.11 $ & $    919 $ & $   7.30 $ & $   0.30 $ & $   3.98 $ & $   5.70 $ & $   3.84 $ & $   1.16 $ & $    106 $ & $  8.1\times 10^{-4}  $ & $   1.16 $ &  n.a. & IIP & 0.003  \\ 
$  2512 $ & $  0.85 $ &      C & $   71.2 $ & $   3.10 $ & $    928 $ & $   7.57 $ & $   0.30 $ & $   3.98 $ & $   5.40 $ & $   3.64 $ & $   1.45 $ & $    152 $ & $  1.3\times 10^{-3} $ & $   1.12 $ &  n.a. & IIP & 0.004 \\ 
$  2512 $ & $  0.80 $ &      C & $   71.3 $ & $   3.09 $ & $    935 $ & $   7.86 $ & $   0.30 $ & $   3.98 $ & $   5.06 $ & $   3.40 $ & $   1.79 $ & $    263 $ & $  1.6\times 10^{-3} $ & $   1.06 $ &  n.a. & IIP & 0.005 \\ 
$  2512 $ & $  0.75 $ &      C & $   71.7 $ & $   3.10 $ & $    929 $ & $   7.55 $ & $   0.30 $ & $   3.98 $ & $   5.48 $ & $   3.70 $ & $   1.37 $ & $   1106 $ & $  4.4\times 10^{-3} $ & $   0.00 $ &  n.a. & IIP & 0.006 \\ 
$  2512 $ & $  0.70 $ &      C & $   71.2 $ & $   3.14 $ & $    902 $ & $   6.64 $ & $   0.30 $ & $   3.98 $ & $   6.71 $ & $   4.54 $ & $   0.14 $ & $   51.9 $ & $ 1.9\times 10^{-4} $ & $   0.00 $ &  n.a. & IIP & 0.004 \\ 
$  2512 $ & $  0.65 $ &      C & $   71.2 $ & $   3.15 $ & $    900 $ & $   6.57 $ & $   0.30 $ & $   3.98 $ & $   6.82 $ & $   4.62 $ & $   0.03 $ & $   5.54 $ & $  2.1\times 10^{-5} $ & $   0.08 $ &  n.a. & IIP & 0.00 \\ 
$  2512 $ & $  0.60 $ &     No & $   71.1 $ & $   3.15 $ & $    898 $ & $   6.54 $ & $   0.30 $ & $   3.98 $ & $   6.84 $ & $   4.63 $ & $   0.009 $ & n.a. & n.a. & n.a. &  n.a.& IIP &  n.a. \\ 
$  2512 $ & $  0.55 $ &     No & $   70.9 $ & $   3.15 $ & $    897 $ & $   6.52 $ & $   0.30 $ & $   3.98 $ & $   6.85 $ & $   4.64 $ & $   0.00 $ & n.a. & n.a. & n.a. &  n.a. & IIP &  n.a. \\ 
$  2512 $ & $  0.50 $ &     No & $   71.1 $ & $   3.15 $ & $    898 $ & $   6.54 $ & $   0.30 $ & $   3.98 $ & $   6.86 $ & $   4.64 $ & $   0.00 $ & n.a. & n.a. & n.a. &  n.a. & IIP & n.a.  \\ 
\hline
$  1995 $ & $  0.90 $ &      C & $ 70.9 $ & $  3.03 $ & $  968 $ & $ 10.8 $ & $  0.30 $ & $  3.98 $ & $  2.56 $ & $  1.68 $ & $  4.30 $ & $ 12.8 $ & $  9.1\times 10^{-4} $ & $ 21.3 $ & n.a. & IIP & $14$ \\
$  1995 $ & $  0.85 $ &      C & $ 71.1 $ & $  3.03 $ & $  968 $ & $ 11.2 $ & $  0.31 $ & $  3.98 $ & $  2.33 $ & $  1.52 $ & $  4.53 $ & $ 11.6 $ & $  1.5\times 10^{-3} $ & $ 21.3 $ & n.a. & IIP & $15$\\
$  1995 $ & $  0.80 $ &      C & $ 70.7 $ & $  3.04 $ & $  963 $ & $ 11.7 $ & $  0.31 $ & $  3.98 $ & $  2.06 $ & $  1.34 $ & $  4.81 $ & $ 10.4 $ & $ 4.0\times 10^{-3} $ & $ 21.2 $ & n.a. & IIP & $16$ \\
$  1995 $ & $  0.75 $ &      C & $ 71.0 $ & $  3.05 $ & $  957 $ & $ 12.3 $ & $  0.31 $ & $  3.98 $ & $  1.76 $ & $  1.13 $ & $  5.10 $ & $  9.38 $ & $ 1.3\times 10^{-2} $ & $ 21.0 $ & n.a. & IIP & $17$ \\
$  1995 $ & $  0.70 $ &      C & $ 71.2 $ & $  3.07 $ & $  944 $ & $ 13.1 $ & $  0.32 $ & $  3.98 $ & $  1.42 $ & $  0.90 $ & $  5.43 $ & $  8.54 $ & $  3.3\times 10^{-2} $ & $ 20.3 $ &n.a. & IIP & $18$ \\
$  1995 $ & $  0.65 $ &      C & $ 70.9 $ & $  3.11 $ & $  918 $ & $ 14.0 $ & $  0.33 $ & $  3.98 $ & $  1.08 $ & $  0.67 $ & $  5.78 $ & $  6.60 $ & $  7.9\times 10^{-2} $ & $ 19.8 $ & n.a. & IIP & $19$ \\
$  1995 $ & $  0.60 $ &      C & $ 70.7 $ & $  3.18 $ & $  879 $ & $ 14.9 $ & $  0.34 $ & $  3.98 $ & $  0.74 $ & $  0.44 $ & $  6.12 $ & $  5.12 $ & $ 0.14 $ & $ 18.7 $ & n.a. & IIL & $21$ \\
$  1995 $ & $  0.55 $ &      C & $ 71.1 $ & $  3.29 $ & $  821 $ & $ 16.0 $ & $  0.38 $ & $  3.98 $ & $  0.45 $ & $  0.25 $ & $  6.40 $ & $  3.40 $ & $ 0.20 $ & $ 17.7 $ & n.a. & IIL & $22$ \\
$  1995 $ & $  0.50 $ & $\text C$ & \multicolumn{4}{l}{Common Envelope Evolution} & $\simgr 0.30$ & $  3.98 $ &   \multicolumn{2}{c}{ $\sim0.2-4.1$}  &  $ 2.72 $ & ? & $ 0.25 $ & $ 16.4 $ & $ 19.6 $ & II  & $\sim9-23$  \\
$  1995 $ & $  0.45 $ & $\text C$ &  \multicolumn{4}{l}{Common Envelope Evolution} & $\simgr 0.30$ & $ 3.98 $ &   \multicolumn{2}{c}{$\sim0.2-4.5$} & $2.33 $ & ? & $ 0.29 $ & $ 14.8 $ & $ 22.5 $ & II  & $\sim8-23$ \\
$  1995 $ & $  0.40 $ &  $\text C$ &  \multicolumn{4}{l}{Common Envelope  Evolution} & $\simgr 0.30$ & $ 3.98 $ &   \multicolumn{2}{c}{$\sim0.2-4.8$}  &  $ 2.08 $ & ? & $ 0.32 $ & $ 13.0 $ & $ 26.0 $ & II  & $\sim7-23$ \\
$  1995 $ & $  0.35 $ &  $\text C$ & \multicolumn{4}{l}{Common Envelope Evolution} & $\simgr 0.30$ & $ 3.98 $ &   \multicolumn{2}{c}{ $\sim0.2-5.0$}  & $ 1.87 $ & ? & $ 0.35 $ & $ 10.7 $ & $ 30.3 $ & II  & $\sim6-23$ \\
$  1995 $ & $  0.30 $ & $\text C$ &   \multicolumn{4}{l}{Common Envelope  Evolution} & $\simgr 0.30$ & $ 3.98 $ &   \multicolumn{2}{c}{ $\sim0.2-5.2$}  & $ 1.70 $ & ? & $ 0.39 $ & $  7.58 $ & $ 36.3 $ & II  & $\sim6-23$ \\
$  1995 $ & $  0.25 $ & $\text C$ & \multicolumn{4}{l}{Common Envelope  Evolution}& $\simgr 0.30$ & $ 3.98 $ &   \multicolumn{2}{c}{ $\sim0.2-5.3$}  &  $ 1.54 $ & ? & $ 0.43 $ & $  3.50 $ & $42.9$  & II & $\sim5-23$ \\
$  1995 $ & $ 0.20 $ & $ \text C $ & \multicolumn{4}{l}{Common Envelope Evolution} &  $\simgr 0.30$ & $ 3.98 $ &   \multicolumn{2}{c}{ $\sim0.2-5.4$}  & $ 1.42 $ & ? & $ 0.48 $ & $  0.74 $ & $52.7$  & II & $\sim5-23$ \\
$  1995 $ & $  0.15 $ &     No & $ 71.1 $ & $  3.15 $ & $  899 $ & $  6.54 $ & $  0.30 $ & $  3.98 $ & $  6.85 $ & $  4.63 $ & $  0.00 $ & n.a. & n.a. &  n.a. & n.a. & IIP & n.a. \\
$  1995 $ & $  0.10 $ &     No & $ 71.0 $ & $  3.15 $ & $  898 $ & $  6.53 $ & $  0.30 $ & $  3.98 $ & $  6.86 $ & $  4.64 $ & $  0.00 $ & n.a. & n.a. & n.a. & n.a. & IIP & n.a.\\
\hline 
$  1413 $ & $  0.90 $ &     BC & $ 66.7 $ & $  3.18 $ & $  854 $ & $ 14.4 $ & $  0.35 $ & $  3.89 $ & $  0.71 $ & $  0.42 $ & $  0.54 $ & $  3.19 $ & $2.9\times 10^{-5} $ & $ 20.6 $ & n.a.  & IIL & $4.4$\\
$  1413 $ & $  0.85 $ &     BC & $ 67.4 $ & $  3.22 $ & $  835 $ & $ 15.1 $ & $  0.36 $ & $  3.88 $ & $  0.59 $ & $  0.33 $ & $  0.39 $ & $  2.68 $ & $2.1\times 10^{-5} $ & $ 18.5 $ & n.a.  & IIL & $3.5$ \\
$  1413 $ & $  0.80 $ &     BC & $ 66.8 $ & $  3.27 $ & $  808 $ & $ 15.3 $ & $  0.37 $ & $  3.87 $ & $  0.48 $ & $  0.26 $ & $  0.26 $ & $  2.06 $ & $1.4\times 10^{-5} $ & $ 15.3 $ & n.a. & IIL & $2.5$ \\
$  1413 $ & $  0.75 $ &     BC & $ 66.7 $ & $  3.30 $ & $  790 $ & $ 15.6 $ & $  0.38 $ & $  3.86 $ & $  0.42 $ & $  0.22 $ & $  0.21 $ & $  1.93 $ &   $1.3\times 10^{-5} $ & $  3.25 $ & n.a.  & IIL & $2.1$ \\
$  1413 $ & $  0.70 $ &     BC & $ 65.2 $ & $  3.40 $ & $  737 $ & $ 15.8 $ & $  0.40 $ & $  3.81 $ & $  0.31 $ & $  0.15 $ & $  0.03 $ & $  0.86 $ & $5.0\times 10^{-6} $ & $  3.44 $ & n.a.  & IIL & $0.33$ \\
$  1413 $ & $  0.65 $ &      B & $ 63.8 $ & $  3.51 $ & $  685 $ & $ 15.9 $ & $  0.43 $ & $  3.76 $ & $  0.25 $ & $  0.10 $ & $  0.00 $ & n.a. & n.a. & n.a. & n.a. & IIb & n.a.\\
$  1413 $ & $ 0.60 $ &  \text{B} &  \multicolumn{4}{l}{Common Envelope Evolution} & $\simgr 0.30$  & $\sim4$ & \multicolumn{2}{c}{$\sim11$} & $0.00$ & n.a. & n.a. & n.a. & $4.58
$ & II & n.a.\\
$    $ & $ \vdots $  &   &  \multicolumn{4}{l}{ }  &  &   &  &  &    &  \\
$  1413 $ & $  0.15$ &  \text{B} &  \multicolumn{4}{l}{Common Envelope Evolution} & $\simgr 0.30$  & $\sim4$ & \multicolumn{2}{c}{$\sim7$} & $0.00$ & n.a. & n.a. & n.a. & $20.0$ & II & n.a.\\
$  1413 $ & $  0.10 $  & \text{C} & \multicolumn{4}{l}{Common Envelope Evolution} &  $\simgr 0.30$ & $  3.98 $ &   \multicolumn{2}{c}{ $\sim0.2-5.7$}  & $ 1.17 $ & ? & $0.91$ & $ 20.2 $ & $31.6$  & II & $ \sim4-23$   \\
\end{tabular}
}
\tablefoot{The models are identified by their initial orbital period (Col.\,1), initial mass ratio (2), and mass transfer case (3). 
Further, we give the properties of the primary stars at core collapse, namely luminosity (4), effective temperature (5), radius (6), spectroscopic luminosity (7), surface helium mass fraction (8), helium core mass (9), envelope mass (10) and the mass of hydrogen in the envelope (11). Column\,12 shows the amount of mass lost via RLOF during stable \case C mass transfer. Column\,13 shows the ratio between the \case C mass transfer rate and wind mass-loss rate at the time of collapse, averaged over the last 100 years. 
Columns\,14 and 15 give the maximum mass transfer rate occurring during \case C RLOF alongside the time before collapse at which it is found. Column\,16 shows the estimated timescale for common-envelope evolution (see Eq.\,\ref{eq:tCE}) when applicable. Column\,17 displays the expected SN type if no CSM interaction would occur, and Column\,18 estimates the energy expected to be thermalized by the CSM interaction  ($E_\text{interact}$, see Eq.\,\ref{eq:E_interaction}), divided by the radiated energy of the SN when the interaction is neglected ($E_\text{LC}\sim 3\times 10^{49}\text{erg}$; \citealt{Dessart2024_widebinaryIIP_from_Ercolino2024}).\\
\tablefoottext{*}{Model did not reach core collapse but terminated after core silicon exhaustion.}
}

\end{table*}

In this section, we describe the evolution of our models
from the ZAMS until the core collapse of the primary star. \Figure{fig:Pq-plot} gives an overview of the 
parameter space covered in this work.
We first discuss the models with an initial mass ratio of $0.95$, and subsequently, we analyze those with smaller initial mass ratios.

\begin{figure*}
   \resizebox{\hsize}{!}{
            \includegraphics[]{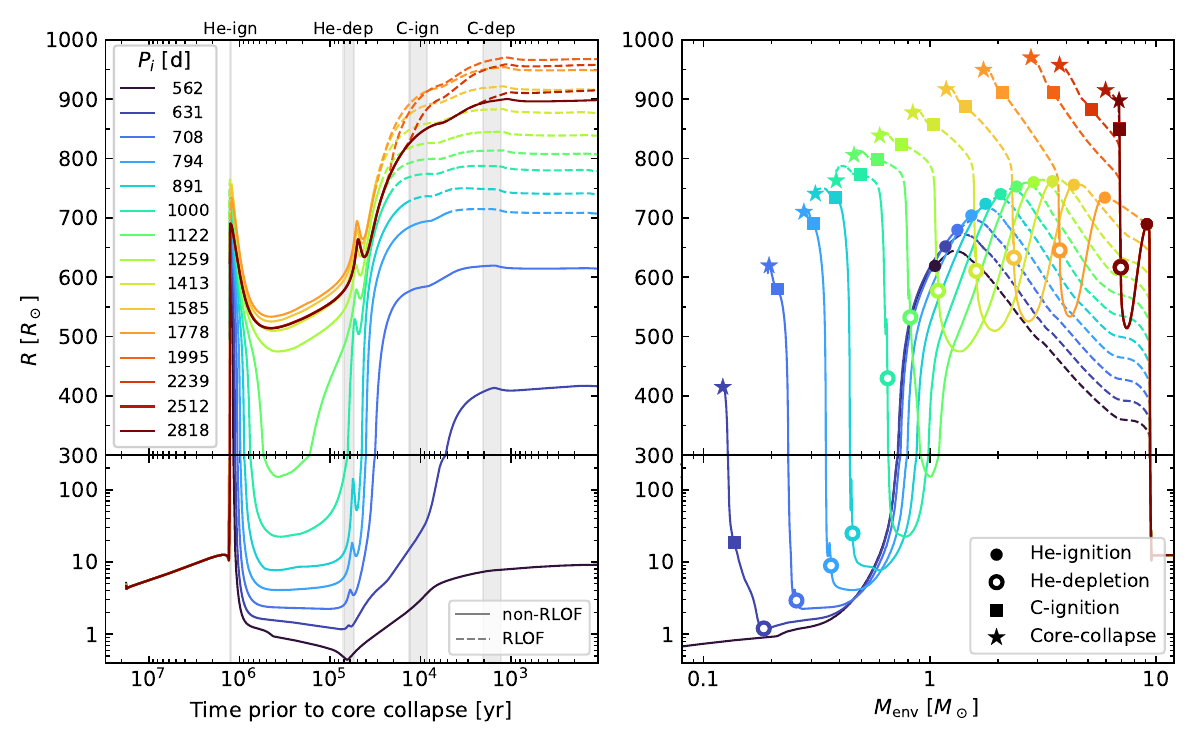}}
    \caption{Evolution of the radius as a function of the time left until core collapse (left) and as a function of the evolving envelope mass (right) of the primary star in models with $\qi=0.95$. Each plot is divided into two parts, with a linear ordinate scale (top) and a logarithmic ordinate scale (bottom). The color coding reflects the initial orbital periods, where dashed lines correspond to phases of RLOF.  Key evolutionary phases are also highlighted by gray vertical bars (left) or markers (right), indicating respectively when core He and C are ignited (-ign) and depleted (-dep).}
    \label{fig:R_vs_t_and_R_vs_Menv}
\end{figure*}

\subsection{Systems with an initial mass ratio of $\qi=0.95$ } \label{sec:models_q0p95_dlogP}
In this section, we consider the binaries where the initial mass of the secondary star is $12.0\Msun$ and the initial orbital period is between $562\days$ and $2818\days$. Of the 15 binary models considered here (cf. Table \ref{table:data}, top part), 14 undergo either mass transfer after core hydrogen exhaustion (\Case B), after core helium exhaustion (\Case C), or both, while the initially widest one does not undergo RLOF anytime during its evolution. 
Since our set of physical parameters is close to those adopted by \citet{Brott2011}, both binary components evolve very similarly to their models until mass transfer occurs. Due to the comparable nuclear timescales of both components, \Case B RLOF happens when the secondary is close to depleting hydrogen in the core ($X_c(\text{H})\simle 0.1$), while \Case C RLOF happens when the secondary is at the bottom of the red giant branch (cf., \Sect{sec:preSN_secondary_star}). 

These binary models undergo stable mass transfer, which reduces the mass of the donor stars on the thermal time scale. 
This is possible even for the widest considered orbits where the donors are RSGs with fully convective envelopes because the orbits widen early on due to the large initial mass ratio.
While this has been found in previous models for \case C mass transfer \citep[][]{ 1993J_Podsiadlowski, 1993J_Woosley}, due to our mass transfer scheme (\Sect{sec:methods}) it occurs here also in wide \case B models \citep[e.g., the \case LB binaries in ][]{Sravan_b}.
The history of mass transfer can be inferred from the mass and radius evolution of the primaries, as shown in \Fig{fig:R_vs_t_and_R_vs_Menv}. Model N2818-0.95 does not undergo RLOF and its primary serves as an analog to a single star.

\subsubsection*{\Case B mass transfer}

\begin{figure}
    \includegraphics[width=\columnwidth]{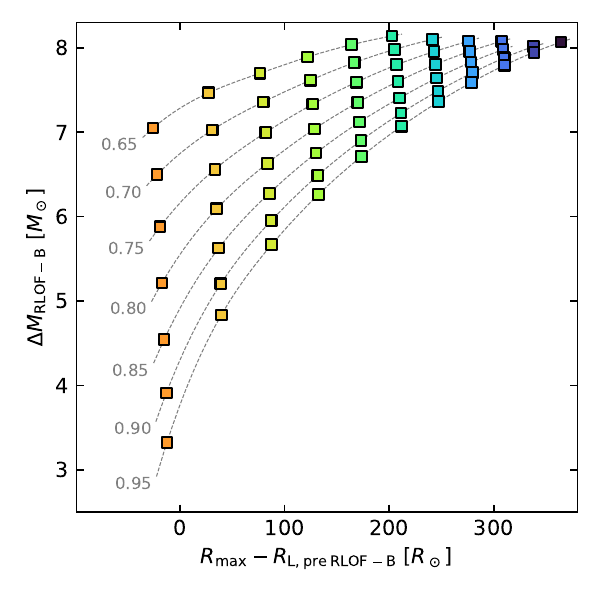}
    \caption{Amount of mass lost during stable \case B mass transfer for all models where this is encountered, as a function of the maximum radius the star would have exhibited in this phase without a companion (i.e., the maximum radius of the model N2818-0.95) minus the Roche lobe radius prior to interaction. Different colors represent different initial orbital periods, as in \Fig{fig:R_vs_t_and_R_vs_Menv}, and the dashed lines connect systems with the same initial mass ratio, whose value is labeled on the left.}
    \label{fig:DeltaM_B}
\end{figure}

\case B mass transfer happens for the systems with   $P_i\leq 1778\days$. The total amount of mass removed from the primary increases monotonically with decreasing orbital period. This trend can be interpreted in terms of the primary star's expansion: the smaller the Roche lobe radius is compared to the maximum extent of the single-star counterpart during this phase, the earlier mass transfer begins, resulting in more mass being transferred. \Figure{fig:DeltaM_B} shows this behavior quantitatively and highlights that for every primary star undergoing \case B RLOF, their Roche lobe radius prior to mass transfer ($R_\text{L, pre\,RLOF-B}$) is smaller than the maximum radius of the non-interacting model in this phase ($R_\text{max}$). A notable exception is the initially widest system undergoing \case B RLOF (BC1778-0.95) in which this difference is slightly negative. In this model, mass transfer was exclusively regulated by the outflow of the extended, optically thin atmosphere above the photospheric radius \citep{Ritter_scheme}, which induced the expansion of the convective envelope of the primary star.

Because the initial mass ratio is close to one, the ratio exceeds one early during mass transfer and then the orbit widens during RLOF. The envelope's expansion following mass removal results in the primary star still overflowing at this stage. This continues until the star ignites helium in the core (\Fig{fig:R_vs_t_and_R_vs_Menv}) as the hydrogen-burning shell becomes less strong, and the envelope responds by contracting.
The star's reaction to mass transfer combined with the orbital evolution explains why the models with initial periods in the range $794 \ldots 1778\days$ expand to higher radii than the maximum radius expected from a single-star model during the same phase (\Fig{fig:R_vs_t_and_R_vs_Menv}). 
The model with the largest radius by the time of helium ignition (BC1259-0.95), with a radius of $750\Rsun$, is not the initially widest \case B system. This result is the combination of two competing effects, as wider systems have initially larger Roche lobe radii due to the larger initial separation, but also shed less mass via mass transfer which drives less expansion of the Roche lobe.

\subsubsection*{Evolution during core-helium burning} \label{sec:coreHeburn}
\begin{figure*}
    \includegraphics[width=0.5\linewidth]{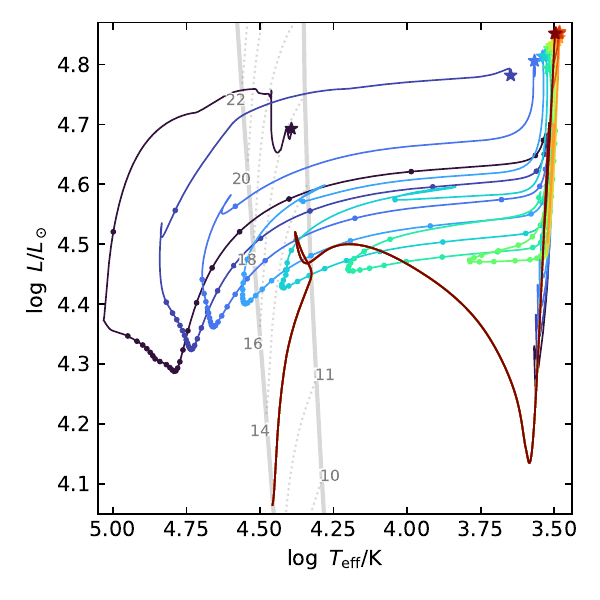}
    \includegraphics[width=0.5\linewidth]{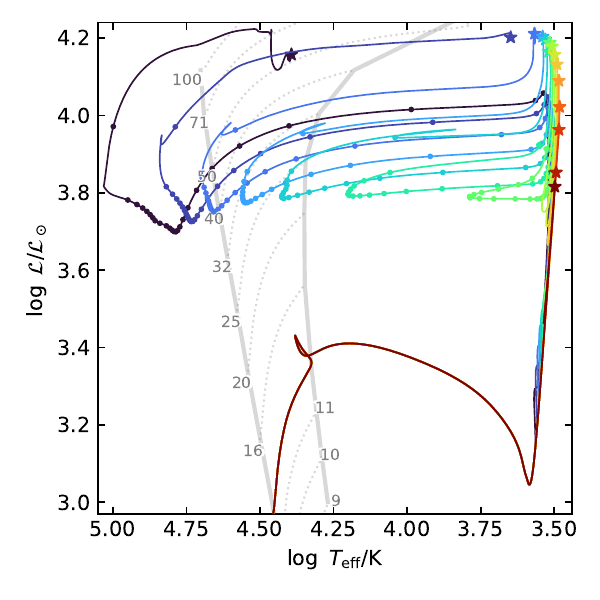}
    \caption{Evolution of the primary stars for our models with $\qi=0.95$ in the HRD (left) and sHRD (right). The markers indicate the position where the primaries reach core collapse, and the color-coding is as in \Fig{fig:R_vs_t_and_R_vs_Menv}. Thick gray lines reflect the ZAMS and terminal-age main-sequence (TAMS) positions of single stars, and thin gray dashed lines show the main-sequence evolution of single stars with initial mass displayed at the edges of the tracks.  For models that leave the RSG regime after core helium ignition, dots are placed on their tracks at a fixed time interval of $50\kyr$.}
    \label{fig:HR}
\end{figure*}

After the ignition of helium, the envelope contracts, causing the end of mass transfer for the systems undergoing \case B RLOF (\Fig{fig:R_vs_t_and_R_vs_Menv}). Following this point, the envelope mass $M_\text{env}$ (defined as the mass of the hydrogen-rich shells, with $X>0.01$; see \Appendix{sec:App_core_envelope} for a discussion on different definitions) is further reduced by wind mass-loss and shell burning. Higher envelope masses following core helium ignition result in stronger winds and core-growth (cf. \Fig{fig:Menv_vs_t}). The only exception to this is the initially tightest model in the set (B562-0.95) which becomes very hot and exhibits stronger winds, leading to the complete loss of the envelope. 

 The growth of the mass of the helium core is proportional to the nuclear luminosity of the hydrogen shell, which is similar in all models by the time of helium ignition in the core. However, during core-helium burning, the more stripped envelopes will exert less pressure on the hydrogen-burning shell, making it less luminous and hindering further core growth.
 The wind strength is affected by the core growth, as less massive cores result in lower luminosities and thus lower mass-loss rates (e.g., \citealt{Vink2001} and \citealt{Nieuw_wind}). However, in the case of the tightest model in this set  (B562-0.95), the envelope is He-enhanced and becomes very hot during the phase of core-helium burning. Consequently, the wind ramps up (cf. \Sect{sec:wind_method}) and removes the remaining envelope in a short timescale. 

In the Hertzsprung-Russell diagram (\Fig{fig:HR}) we see that the luminosity is similar for all models, as it is mainly set by the helium core mass, which varies by at most $0.56\Msun$ between the initially widest and tightest systems (\Fig{fig:Menv_vs_t}). 
The effective temperature however differs more significantly. A blue-ward motion across the HR diagram occurs for models in which $M_\mathrm{env}\lesssim 0.7M_\text{He-core}$ (i.e., those with $P_\text i<1122\days$), and is more extreme for smaller envelope masses.
Some models eventually cross the main-sequence, and the bluest region of these loops is where the stars spend the majority of the time during core helium burning (up to $\sim10^5\text{yr}$, cf. \Fig{fig:HR}). 

Such stars could be mistaken for main-sequence stars just based on their colors and magnitudes. 
They have similar radii as main-sequence stars of the same luminosity, but far less mass, and thus a lower surface gravity $g$. Furthermore, they would be strongly nitrogen and helium enriched (see below).
In the spectroscopic HR diagram (\Fig{fig:HR}),
they share the same location as very massive main-sequence stars. 

The \case B models with $M_\mathrm{env}> 0.7M_\text{He-core}$, as well as those that have not yet undergone interaction, perform a much smaller loop in the HR diagram and stay close to the Hayashi line. In particular, the \case B systems that do not move blue-wards are redder than the non-\case B models, with up to 100\,K lower surface temperatures.     

Once helium is depleted in the core, the models expand again, following the ignition of the helium-burning shell. The partially stripped models evolve back towards the Hayashi line to become RSGs once again. The fully stripped \model{B}{562}{0.95} does not expand above $12\Rsun$. \model{B}{631}{0.95} has retained a H-rich envelope of $0.12\Msun$ and expands to $440\Rsun$ to explode as a yellow supergiant. 

\subsubsection*{\Case C mass transfer}\label{sec:caseC_q0p95_dlogP}

\begin{figure}
    \includegraphics[width=\linewidth]{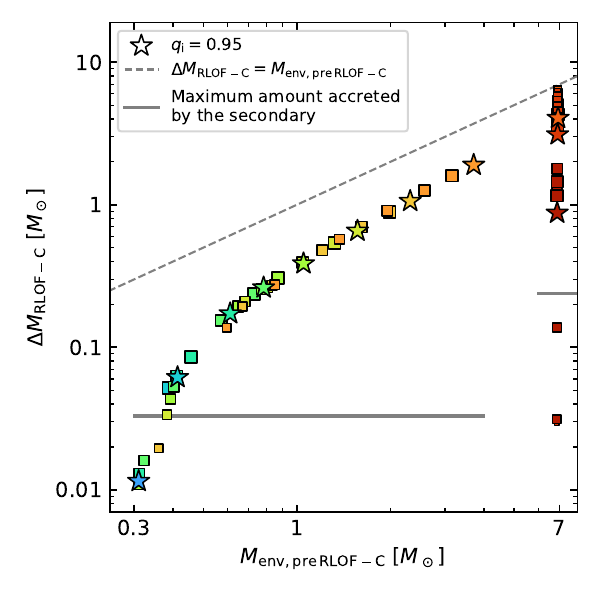}
    \caption{ Amount of mass lost by the primary star during stable \case C mass transfer as a function of the envelope mass prior to \Case C RLOF. Models with different initial orbital periods are shown in different colors, as in \Fig{fig:R_vs_t_and_R_vs_Menv}. Models with $\qi=0.95$ are marked with a star, and models with smaller initial mass ratios are shown with smaller square markers. The gray horizontal lines mark the maximum mass that the secondaries accrete during the first (right) or second (left) phase of mass transfer.}
    \label{fig:DeltaM_C}
\end{figure}

\begin{figure}
    \includegraphics[width=\columnwidth]{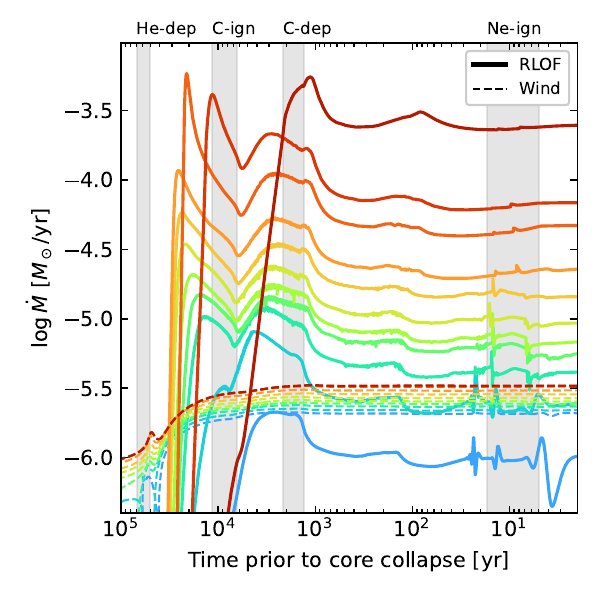}
    \caption{Mass-transfer rate and wind mass-loss rate as a function of time prior to collapse for the models with $\qi = 0.95$. The color coding and vertical gray bars are the same as in \Fig{fig:R_vs_t_and_R_vs_Menv}. The phase of neon ignition is also highlighted (Ne-ign).}
    \label{fig:MdotC_q0p95}
\end{figure}

The envelope expansion following core helium depletion leads to \Case C RLOF for all \Case B models with $P_\text i\geq 794\days$, as well as for those models that have not yet undergone any interaction with $P_\text i \leq 2512\days$. For the initially tighter systems which underwent \case B RLOF, the expansion is too small to reach their post-\case B Roche lobe radius  (cf. the radius at the moment of helium ignition and the expansion following core helium depletion in \Fig{fig:R_vs_t_and_R_vs_Menv}) and will thus reach core collapse without undergoing further mass transfer. For Models C1995-0.95, C2239-0.95, and C2512-0.95, this is the first and only mass transfer event. Their mass ratio exceeds unity before the onset of mass transfer due to wind mass-loss.

All primaries undergoing \case C RLOF, and also our \Case BC models, develop again an extended convective envelope, which keeps the star from detaching once it fills its Roche lobe. As such it is expected that the systems undergoing \case C RLOF will remain in this predicament until core collapse. 

The amount of mass lost in this phase scales with the envelope mass prior to interaction  (\Fig{fig:DeltaM_C}), in spite of the fact that the Roche lobe radii depend non-monotonically on the envelope mass at this time. Exceptions to this trend can be found in the low- and high-envelope mass regimes, as these models begin to overflow later and thus have less time to transfer mass (see the tracks of Models BC794-0.95, BC891-0.95 and C2512-0.95 in \Fig{fig:MdotC_q0p95}). The system that transfers the most mass in this phase is C1995-0.95, which is the tightest system undergoing exclusively \case C RLOF, and it sheds more than half of its envelope. At the other end of the spectrum, primaries that retained $M_\mathrm{env}\simle0.30\Msun$ (or similarly a hydrogen mass $M_\text H\simgr 0.10\Msun$) do not interact again.   

For the majority of the models undergoing \case C or \case BC mass transfer, the time dependence of the mass transfer rate shows two maxima, corresponding to the onset of RLOF (we call it \case C-a), and to the phase of core-carbon burning (\case C-b; see \Fig{fig:MdotC_q0p95}).
\Figure{fig:MdotC_q0p95} shows that the final mass transfer rate is a strong function of the initial orbital period of our models, with values largely exceeding the stellar wind mass-loss rates of the primary stars. 

\begin{figure}
    \includegraphics[width=\columnwidth]{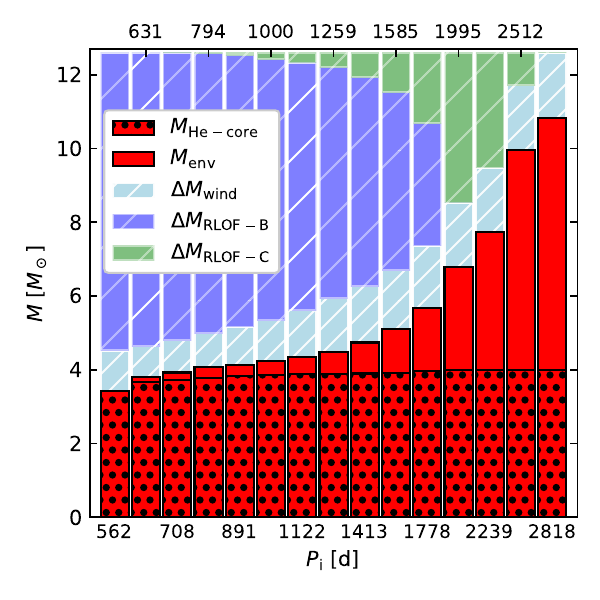}
    \caption{Stacked-bar chart showing how much of the primary star's initial mass of 12.6$\mso$ is left at core collapse (red), and how much is lost
    (white hatching), as a function of the initial binary
    orbital period for our systems with an initial mass ratio of $\qi=0.95$. The 
    helium core mass (red filling, black dots) is distinguished from the mass of the remaining H-rich envelope (red filling, no hatching) with a black line. The three colored white-hatched bars represent the mass lost via wind (azure), \Case B RLOF (blue), and \Case C RLOF (green).}
    \label{fig:Mbar}
\end{figure}

The winds, \case B mass transfer, and \case C mass transfer remove mass with different trends with increasing initial orbital periods (cf. \Fig{fig:Mbar}). Combined, these three result in a monotonic increase in the final envelope mass as a function of the initial orbital period, as can be seen in \Fig{fig:Pq-plot}. 

\subsubsection*{Changes in surface composition} \label{sec:FDU}
As our models include rotational mixing, nuclear-processed material may already be transported to the envelope during the main-sequence. However, this effect is not strong given the initial rotation rate of only $20\%$ critical. By the time of core-hydrogen depletion, the surface abundance of $^{14}\text{N}$ is enhanced by $\sim 17\%$ \citep[cf.,][]{Brott2011}. 

Thereafter, while the stars climb the RSG branch, the so-called first dredge-up occurs. As the star is expanding during this phase, RLOF may occur before the convective region reaches the base of the envelope. If this is the case, it is expected that the amount of enrichment of He and CNO-processed material will be affected, as there will be a smaller envelope to dredge material into.

The models that undergo \case B RLOF show a post-RLOF surface helium abundance of up to $Y_\text s=0.56$ in the most-stripped model, and a surface mass fraction ratio N/C of at most $300$. In comparison, these quantities are $0.30$ and $28$ respectively for the ones not undergoing \case B RLOF after the end of the first dredge up. These abundance patterns remain mostly unaltered during the following evolution unless the system exhibits a further phase of mass transfer. In this case, more He- and CNO-processed material will be exposed, depending on the degree of stripping.

\subsection{Binary models with $\qi \leq 0.95$}

Initially, given a fixed orbital period, systems with lower secondary masses are tighter, while the primary's Roche lobe radius \citep{Eggleton_RL} is instead larger. 
This implies that RLOF will start later on in the evolution and that the orbits will shrink more. As our models with $\qi=0.95$ (cf., \Sect{sec:models_q0p95_dlogP}), all our models with smaller mass ratios develop deep convective envelopes by the time mass transfer starts. This means that once mass transfer ensues, a potentially unstable phase of mass transfer may develop, as the expanding radius of the primary star will meet a strongly shrinking Roche lobe, which may lead to unstable mass transfer. As such it is expected that only models with initial mass ratio above a certain threshold may undergo stable mass transfer.

Below, we discuss in more detail three sets of models, with initial orbital periods of $P_\text i=1413$, $1995$ and $2512\days$, and varying initial mass ratio (cf. \Fig{fig:Pq-plot} and \Tab{table:data}).

\subsubsection*{Systems with $P_{\text i}=1413\days$}
All models except the one with the lowest considered mass ratio (C1413-0.1) undergo \Case B mass transfer. 
\Case B mass transfer has an increasing maximum mass transfer rate with decreasing $\qi$, as well as total mass transferred (cf. \Fig{fig:DeltaM_B}). By the time mass transfer stars, wind mass-loss from the primary during the main-sequence is too weak to reverse the mass ratio, so the mass transfer rate always exhibits a strong initial phase, during which the majority of the mass is transferred.

Models with $\qi\leq0.60$ experience unstable mass transfer. This is compatible with \cite{Pavlovskii_Ivanova_2015_MT_from_Giants}, in which they found the critical mass ratio of $0.45-0.65$. As the orbits are tighter and the envelopes are more bound in this phase compared to the models at  $P_\text i =1995\days$ (see below), these binaries are expected to merge. We assume that the majority of the envelope remains bound \citep[as in the hydrodynamical simulations in][]{Betelgeuse_merger}, and the systems merge. As this happens at core helium ignition, these binaries would not contribute to binary-induced interacting SNe. The model with $\qi=0.65$ does not undergo unstable mass transfer, but its hydrogen-rich envelope does not expand significantly following core-helium depletion. This model will thus not undergo a second phase of mass transfer, and it will not result in an interacting SN.

\case C mass transfer follows for the models that retained an envelope $M_\mathrm{env}> 0.3 \Msun$ following \case B mass transfer, which is found for systems with $\qi>0.65$. The mass shed during \case C mass transfer scales with increasing pre-interaction envelope mass (cf. \Fig{fig:DeltaM_C}), which itself scales with $\qi$. Also as $\qi$ decreases, the time between the beginning of \case C and core collapse decreases (\Fig{fig:MdotC_dq_1413_1995}), further decreasing the amount of mass that will be shed. Models with $\qi>0.75$ clearly exhibit both \case C-a and \case C-b mass transfer phases, while the model with $\qi=0.70$ only exhibits the \case C-b phase due to the later start of the mass transfer.

Finally, the Roche lobe radius of \Model{C}{1413}{0.10} is wide enough to only trigger \case C mass transfer, but given the extreme initial mass ratio, mass transfer quickly becomes unstable and is expected to lead to a common envelope phase. 
As the low mass of the secondary results in a lower orbital potential energy, this system cannot unbind its envelope completely. 
However, the in-spiral is expected to be still ongoing 
by the time the primary star undergoes core collapse.

\subsubsection*{Systems with $P_{\rm i}=1995\days$}\label{sec:dq3.30}
\begin{figure*}
    \includegraphics[width=0.5\linewidth]{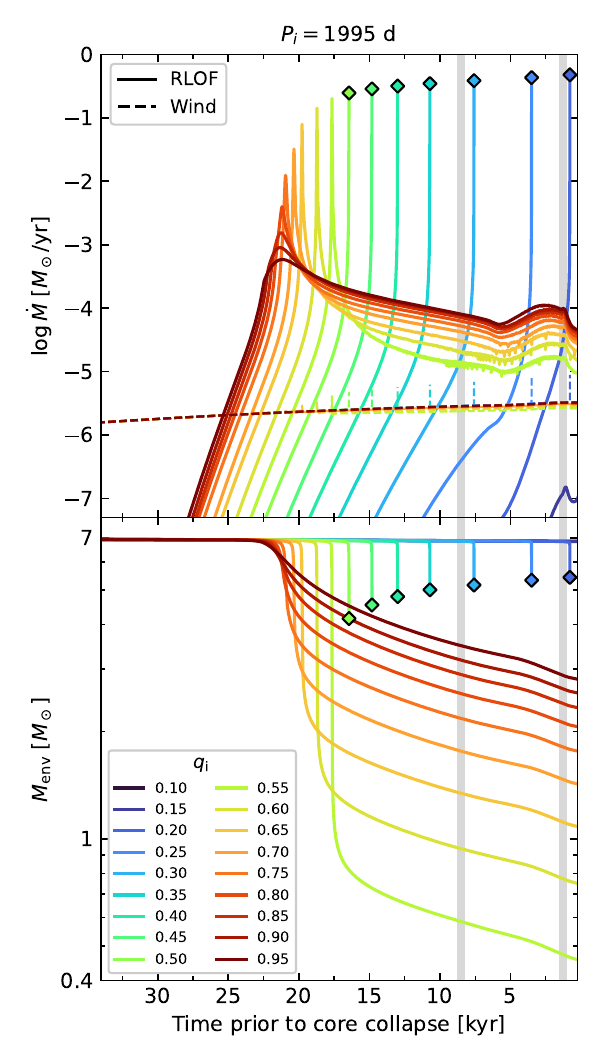}
    \includegraphics[width=0.5\linewidth]{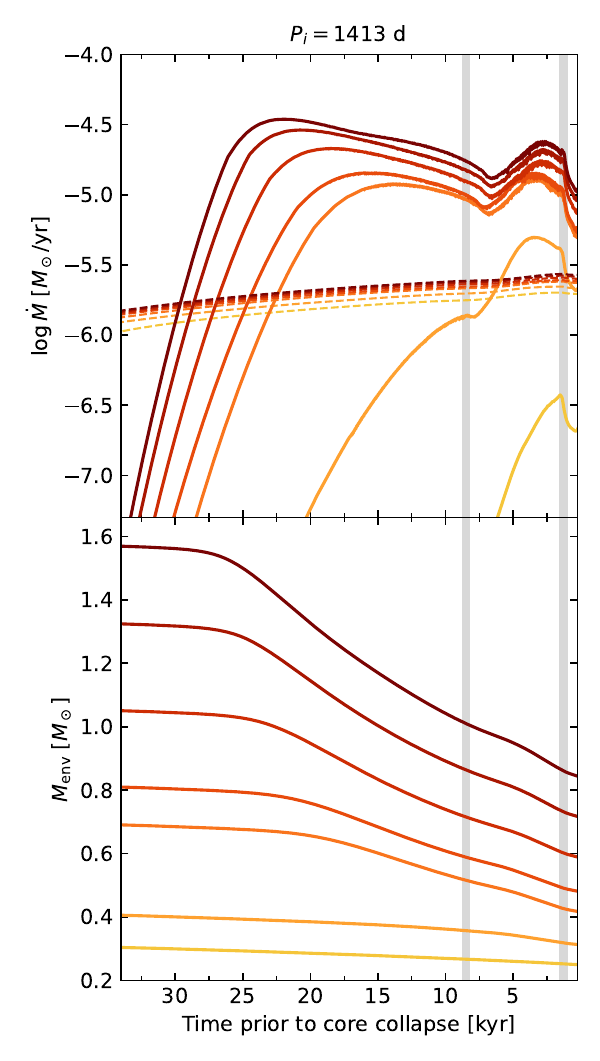}
\caption{Mass-loss rate (top) and envelope mass (bottom) as a function of time prior to collapse in the models with $P_\text i = 1995\days$ (left)  and $P_\text i =1413\days$ (right), and with different $\qi$ shown in different colors (see legend). The vertical gray bars highlight, from left to right, the moment of ignition and depletion of carbon in the core. The markers represent the start of unstable mass transfer. \Model{C}{1413}{0.10} is not shown, but its behavior is similar to \Model{C}{1995}{0.20}.}
    \label{fig:MdotC_dq_1413_1995}
\end{figure*}

The systems in this set with $\qi>0.15$ undergo RLOF and do so only after core helium exhaustion, which occurs roughly $\sim55\kyr$ prior to collapse. Mass transfer starts later for systems with smaller initial mass ratios (cf., \Fig{fig:MdotC_dq_1413_1995}). Despite having less time to transfer mass prior to collapse, the systems with smaller $\qi$ undergoing stable mass transfer lose more mass overall, because their binary orbit shrinks faster.

During core helium burning, the primaries lose about $1.5\mso$ due to their RSG wind.
This allows the models at $\qi=0.95$ and $0.90$ to already invert the mass ratio before \case C mass transfer starts.
In contrast, the models with $\qi\leq0.50$ develop unstable mass transfer after having transferred less than half of the envelope mass (cf. \Tab{table:data}) with mass transfer rates that exceed the thermal-timescale mass transfer rate $\dot M_\text{th}=M_\text{env}/\tau_\text{KH}\simeq 0.25\msoy$. 

The time evolution of the mass transfer rate of these systems shows again two peaks (see \Sect{sec:caseC_q0p95_dlogP}), with an initial burst (\case C-a) followed by a second maximum after carbon is ignited in the core (\case C-b). 
The mass transfer rate during \case C-a, where most of the mass is transferred, reaches a higher and sharper peak for decreasing $\qi$ (see \Fig{fig:MdotC_dq_1413_1995}) since more mass needs to be transferred to invert the mass ratio. The mass transfer rate drops once the orbit widens enough for the Roche lobe to catch up with the primary's radius. \Case C-b follows once carbon is ignited in the core, undergoing a similar time evolution as described in \Sect{sec:caseC_q0p95_dlogP}. Here the mass transfer rate correlates with the envelope mass retained following \case C-a.

The systems with $\qi<0.55$ undergo unstable-mass transfer and enter a CE phase. The envelope binding energy is sufficiently small such that it could be ejected. However, the onset of the
unstable mass transfer occurs close to core collapse (cf., the diamonds in \Fig{fig:MdotC_dq_1413_1995}), the more so the more extreme
the initial mass ratio. In \Model{C}{1995}{0.20}, this happens only near core-carbon depletion, some $2\,000\years$ before core collapse.
 
\subsubsection*{Systems with $P_{\text i}=2512\days$}
\begin{figure}
    \includegraphics[width=1.0\columnwidth]{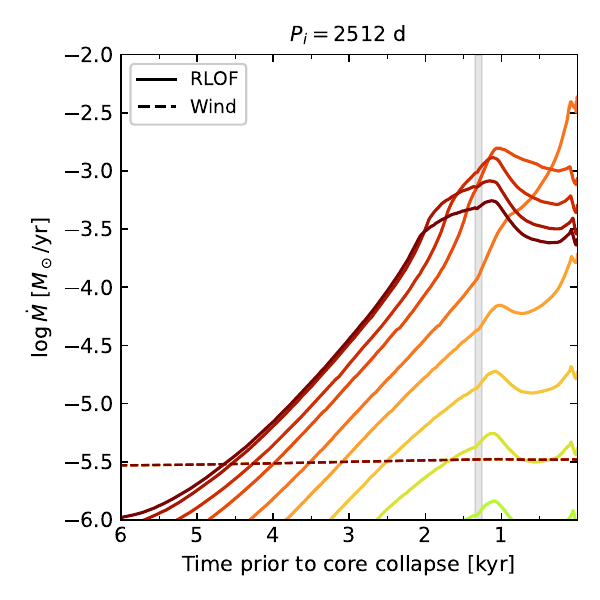}
\caption{Mass-loss rate as a function of time prior to collapse for the models with $P_\text i = 2412$ d, with different $\qi$ shown in different colors as in \Fig{fig:MdotC_dq_1413_1995}. The vertical gray bar highlights the moment of core carbon depletion.}
    \label{fig:MdotC_dq_2512}
\end{figure}

The models in this set begin mass transfer just $5\kyr$ prior to core collapse. During this time, the mass transfer is not enough to invert the mass ratio, and thus all models explode while the orbit is still tightening. 

For models with $\qi\geq0.80$, the mass transfer is enabled by the extended RSG atmosphere \citep{Ritter_scheme}.
The mass ratio is already quite close to unity, and thus the Roche lobe radii can keep up with the already expanded envelope (cf. \Tab{table:data}). The mass transfer rate shows a peak near the time of core-carbon depletion (as in \case C-b), after which the envelope slightly contracts and limits the increase in the mass-transfer rate (cf. \Fig{fig:MdotC_dq_2512}). The peak of mass transfer gets as strong as $10^{-3}\msoy$, and in total up to $1.8\Msun$ are removed from the envelope by the time of core collapse. 

The models with $\qi= 0.70$ and $0.65$ also transfer mass only through their extended envelopes. As the mass transfer started very late, the envelope does not have time to significantly expand and, even as the Roche lobe radius contracts, the mass-transfer rate stalls once the primary reaches core carbon depletion.

In the model with $\qi=0.75$, differently from the models with lower $\qi$, the rate of mass transfer is high enough to induce a significant expansion of the envelope, even as the star undergoes core-carbon depletion. The Roche lobe radius also shrinks more significantly than in the models at higher $\qi$. 
This model achieves the largest mass transfer rate at core collapse, with a value of $3\times10^{-3}\msoy$.

\subsection{Trends in the considered parameter space}

\begin{figure*}
    \resizebox{\hsize}{!}{\includegraphics{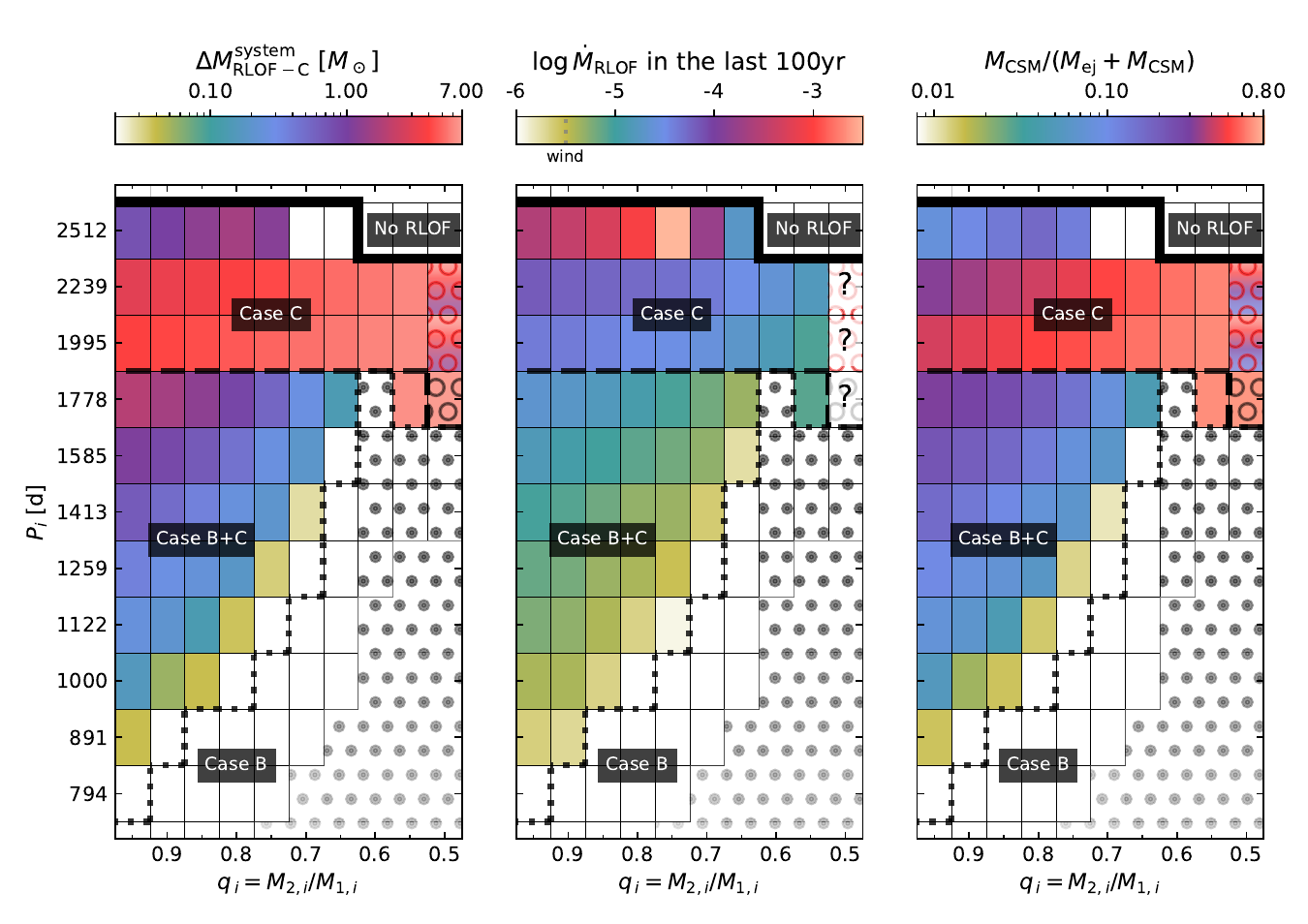}}
    \caption{$P_\text i$--$\qi$ diagrams for our models as in \Fig{fig:Pq-plot}, focusing on the models with $q_{\rm i} \geq 0.5$, where the color-coding now denotes different properties. From left to right, they show the mass lost by the system (i.e., not accreted by the secondary) during \case C mass transfer
    ($\Delta M_\text{RLOF-C}^\mathrm{system}$), the average mass-loss rate due to RLOF during the last 100 years prior to the end of the simulation ($\dot M_\text{RLOF}$), and the fraction of the SN kinetic energy that is expected to be converted into radiation by the interaction ($M_\text{CSM}/(M_\text{ej}+M_\text{CSM})$, see \Sect{sec:interaction_with_RLOF_CSM}). The hatching, as well as contour lines, are the same as in \Fig{fig:Pq-plot}.
    Since the pre-SN envelope properties of the two longest-period Case\,C models with $q_{\rm i}=0.5$ are rather unconstrained, we give the corresponding squares in the first and third panels a range of background colors indicating the range of possibilities. In the middle panel, we use a question mark for the three models which are expected to explode during
    or after a common envelope phase.}
    \label{fig:Pq-plot_zoom}
\end{figure*}
So far we have considered the main features of four subsets of the investigated parameter space. Here we briefly discuss the general trends which appear in the entire explored parameter space (cf. Figs.\,\ref{fig:Pq-plot} and \ref{fig:Pq-plot_zoom}). A comparison between some of these results and previous studies is available in \Appendix{sec:APP_compare}.

\subsubsection*{\Case B systems}
\Case B mass transfer is stable for all the simulated models with $\qi\geq 0.65$, and the overall mass being shed from the primary increases with decreasing $P_\text i$ and decreasing $\qi$ (cf. \Fig{fig:DeltaM_B}). After \case B RLOF, the winds remove $\simle1\Msun$ of the envelope, which allows those that retained less than that to completely shed their envelopes during core-helium burning. Those that have instead retained $M_\text{env}<0.3\Msun$ (or $M_\text H<0.1\Msun$) after core helium depletion do not expand enough to fill their Roche lobe again. Indeed, even a small envelope of $\sim\,0.1\Msun$ is enough for the star to already exhibit an extended envelope of more than $200\Rsun$, which can reach up to $600\Rsun$ for $M_\text{env}=0.3\Msun$ by the time of collapse (cf. \Fig{fig:Rad_Teff_vs_Menv_preCC}).

Systems undergoing unstable  \case B mass transfer are those with $\qi\leq0.60$, all of which are expected to merge following the phase of CE evolution.  We will not provide quantitative estimates of the post-merger structure, as this is outside the scope of this work. However, as these models are similar to the ones calculated in \cite{Betelgeuse_merger}, we expect them to share a similar evolution, in which the post-merger product retains a relatively massive envelope.

\subsubsection*{\Case B+C systems}
All systems that have undergone \case B RLOF and retained an envelope $M_\text{env}\simgr 0.3\Msun$ after core helium exhaustion expand enough to fill their Roche lobes again, and will transfer mass proportionally to their envelope masses prior to \case C RLOF (cf. \Fig{fig:DeltaM_C}). As \case B mass transfer has already reversed the mass ratio, this second mass transfer is always stable. These systems do not detach before collapse, and show higher final mass transfer rates for larger final envelope mass, with values of up to $10^{-4}\msoy$ (cf., \Fig{fig:Pq-plot_zoom}, central panel). The systems that retain the higher envelope masses at the time of collapse are those with larger initial period and initial mass ratio, as also shown in \cite{Ouchi2017_IIb_RSG_progenitors}. 

\subsubsection*{\Case C systems with stable mass transfer}
For the widest interacting systems, \Case C mass transfer is stable for $\qi\geq0.55$. Like for systems undergoing \case B RLOF, here the amount of mass transferred is higher in systems with initially lower $P_\text i$ and lower $\qi$ (cf. \Fig{fig:DeltaM_C}). Also in these systems, mass transfer will continue until the moment of core collapse \citep[in qualitative agreement with the recent works of][see also \App{sec:App_comp_caseC}]{Ouchi2017_IIb_RSG_progenitors, Matsuoka_Sawada_BinaryInteraction_IIP_Progenitors}, at which time the mass transfer rates can exceed values of several times $10^{-3}\msoy$ (cf., \Model{C}{2512}{0.75} and \Fig{fig:Pq-plot_zoom}, central panel).

\subsubsection*{\Case C Systems with unstable mass transfer}\label{sec:unstable_caseC}
Systems with $\qi\leq 0.50$ undergo unstable \case C mass transfer. In contrast to the \Case B systems, the orbital energy of the smallest possible
orbit during a common envelope evolution exceeds the envelope binding energy in most of them, except for the \case C systems with $P_\text i \leq 1778\days$ and $\qi\leq 0.20$.

The CE evolution during \case C mass transfer has a limited timescale to proceed, as the primary's core has a remaining lifetime of less than $\sim 20\kyr$. Systems with higher $P_\text i$ and lower $\qi$  (i.e., those with larger Roche lobe radii) start RLOF later, reducing the time available prior to collapse, while at the same time $\tau_\text{CE}$ increases for systems in which the secondary star is less massive (cf. Table \ref{table:data}). This means that the initially tightest \case C systems which undergo CE evolution (i.e., those with lower $P_\text i$ and higher $\qi$, e.g., \Model{C}{1778}{0.50}; see \Fig{fig:Pq-plot}) may be able to eject the CE just prior to collapse. 

On the other hand, the common envelope phase would still be ongoing at the time of the SN, independent of whether there is enough energy to eject the envelope (e.g., \Model{C}{2239}{0.40}) or not (as in \Model{C}{1585}{0.20}). 
While our estimates about the common envelope evolution are very simplistic, the qualitatively different situations at the time of SN may well have a reflection in reality.
We discuss the possible consequences for the 
SN display in \Sect{sec:Supernova_Types}.

\subsubsection*{Outliers}\label{sec:outliers}
Five models deviate from the behavior so far described, i.e., Models BC1778-0.55, C2239-0.35, C2512-0.65, C2512-0.70, C2512-0.75. These models start mass transfer shortly before their radius starts shrinking.

\Model{BC}{1778}{0.55} is located close to the boundary between unstable \case B mass transfer and no \case B interaction. In this system, RLOF starts right when the star begins helium burning, and detachment occurs after only $0.02\Msun$ of material is transferred during \Case B.
It therefore behaves more like a pure \Case C binary.

The other four models fill their Roche lobes less than $7\,000\years$ prior to collapse, and could only manage to transfer less than $1\Msun$ before core collapse. 

Some models not classified as undergoing RLOF also pertain to this category as they removed less than $0.01\Msun$ prior to collapse (e.g., \Model{C}{2512}{0.60}, which only transfers $0.009\Msun$ prior to collapse).

\section{Expected supernovae disregarding the CSM}\label{sec:SNE_noCSM}

\begin{figure*}
    \includegraphics[width=0.5\linewidth]{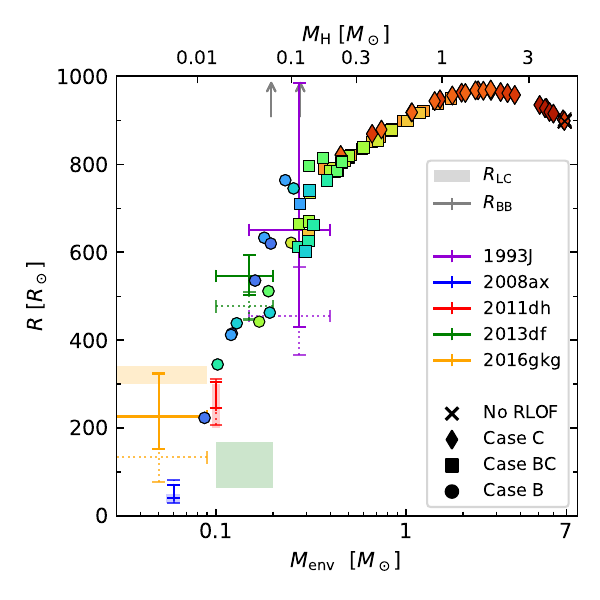}
    \includegraphics[width=0.5\linewidth]{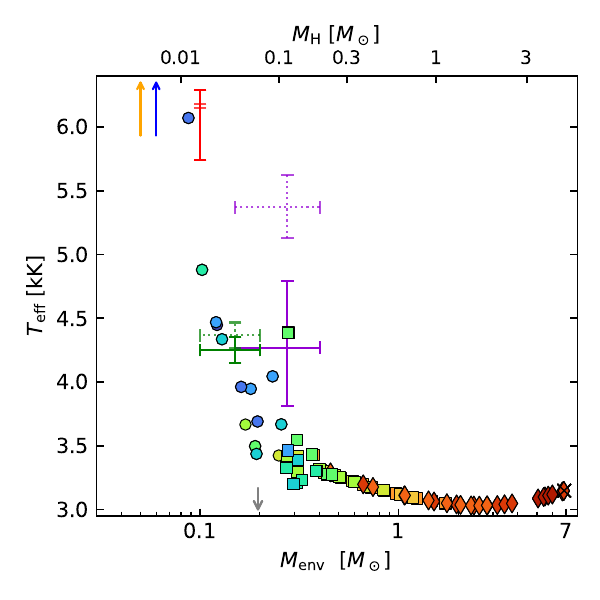}
    \caption{Radius (left) and effective temperature (right) of the primary stars at core collapse as a function of their envelope mass, for models that undergo stable mass transfer and that are not fully stripped (filled markers). Different colors are defined as in \Fig{fig:R_vs_t_and_R_vs_Menv}, and different symbols indicate their mass transfer history (crosses for models not undergoing mass transfer, diamonds for \case C models, squares for \case BC and circles for \case B). Grey arrows indicate models with values outside of the bounds of the plot. The total mass of hydrogen in the models is displayed as a secondary axis on top.
    Estimated progenitor properties of several observed \Type IIb SNe are also shown (cf. \Tab{tab:SNe}) with error bars. Solid-line error bars refer to several references (see \Tab{tab:SNe}) while those in dotted lines are taken from \cite{Gilkis_fits}. The progenitor radius estimates from light-curve fitting are shown as shaded boxes. }
    \label{fig:Rad_Teff_vs_Menv_preCC}
\end{figure*}
\begin{table*}[]
    \caption{Data for observed SN-IIb progenitor stars, similarly to what has been reported in \cite{Yoon_IIb_Ib} and \cite{Sravan_b_preexp}. }
    \centering
    \resizebox{\textwidth}{!}{
\begin{tabular}{c|cc|ccc|ccc}
    & \multicolumn{2}{c}{From light-curves} &  \multicolumn{6}{c}{From pre-explosion imaging}             \\
Name                      & $M^\text{ext}_\text{env}$            & $R_\text{LC}$           & $L$                          & $ T_\text{eff}$                & $R_\text{BB}$              & $L$                       & $ T_\text{eff}$           & $R_\text{BB}$             \\
                          &    $\Msun$                & $\Rsun$                 & $\text k\lso$                &    kK                          & $\Rsun$                    & $\text k\lso$             &  kK                       & $\Rsun$                   \\
\hline
\multirow{1}{*}{1993J}    & $^{a,b}\  0.15-0.40$      &       $/$               & $^c\ 126^{+125}_{- 62}$      & $^c \ 4.27^{+ 0.52}_{- 0.46}$  & $651^{+335}_{-221}$        &  $155^{+ 74}_{- 50}$      & $ 5.37^{+ 0.25}_{- 0.24}$ & $455^{+110}_{- 89}$       \\[0.15cm]
\multirow{1}{*}{2008ax}   & $ ^{d,*}\ (0.06) $          & $ ^d\ 30-50 $           & $^d\ 49.0^{+151}_{-23.7}$    & $^d\ 11.2^{+8.2}_{-3.6}$       & $^d\ 58.9^{+11.1}_{-18.9}$ & $ 89.1^{+ 58.8}_{- 35.4}$ & $14.5^{+ 3.7}_{- 3.0}$    & $ 47.7^{+ 33.0}_{- 19.5}$ \\[0.15cm]
\multirow{1}{*}{2011dh}   & $^f\ 0.10$                & $^{f,g} \ 200-300 $     & $^e\ 88.1^{+  9.6}_{-  8.7}$ & $^e \  6.01^{+ 0.28}_{- 0.27}$ & $274^{+ 31}_{- 28}$        & $ 83.2^{+ 42.7}_{- 28.2}$ & $ 6.17\pm0.01$            & $253^{+ 58}_{- 47}$       \\[0.15cm]
\multirow{1}{*}{2013df}   & $^{i,\dagger}\ 0.10-0.20$ & $^{i,\dagger}\ 64-169 $ & $^h\ 87.1^{+ 12.9}_{- 11.2}$ & $^h\ 4.25\pm0.10$              & $546^{+ 47}_{- 44}$        & $ 74.1^{+  7.2}_{-  6.5}$ & $ 4.37\pm0.10$            & $477^{+ 32}_{- 30}$       \\[0.15cm]
\multirow{1}{*}{2016gkg}  & $^j\ 0.01-0.09$           & $^j\ 300-340 $          & $^j\ 126^{+60}_{-45}$        & $^j\ 7.25^{+ 0.90}_{- 0.85}$   & $^j\ 226^{+98}_{-73}$      & $ 191^{+ 85}_{- 59}$      &  $ 10.5^{+ 3.0}_{- 2.3}$  & $ 133^{+ 95}_{- 55}$      \\
   \end{tabular}}
   \tablefoot{Column 2 contains the estimated mass of the extended envelope, and column 3 shows the estimated progenitor radius, both as derived from light-curve modeling. Columns 4-6 show the estimated luminosity, effective temperature, and progenitor radius from pre-explosion imaging. Columns 7-9 summarize data for the pre-explosion properties of each progenitor as elaborated by \cite{Gilkis_fits}, containing respectively the same quantity as in columns 4-6. \\
   \tablefoottext{*}{This value comes from one example binary evolution model which reproduces the $L$ and $T_\text{eff}$ of the pre-explosion image at the end of its evolution \citep{Folatelli_2008ax}}. \tablefoottext{$\dagger$}{The source of these values uses the semi-analytical method presented in \cite{sa_Menv_SN_NP14}.}}
   \tablebib{
    (\SN{1993J}) $^a$\citealt{1993J_Woosley}, $^b$\citealt{Houck_Fransson_1993J} and $^c$\citealt{Maund};
    (\SN{2008ax}) $^d$\citealt{Folatelli_2008ax}; 
    (\SN{2011dh}) $^e$\citealt{Maund_2011dh}, $^f$\citealt{Bersten_2011dh} and $^g$\citealt{Ergon_2011dh};    (\SN{2013df}) $^h$\citealt{VanDyk_2013df} and $^i$\citealt{MG_2013df}; (\SN{2016gkg}) $^j$\citealt{Bersten_2016gkg}. }
    \label{tab:SNe}
\end{table*}

In this section, we provide a tentative association between the different types of models discussed above and corresponding observed SN types. For this, we first ignore the CSM. In this case, the SN properties depend on the progenitor envelope properties at the time of core collapse. In \Sect{sec:SupernovaCSM}, we assess the CSM properties of our different types of models, the associated properties of the ejecta--CSM interaction, and how the associated SN type is altered. 

Of key importance for the SN display is the progenitor core mass \citep{Luc2011, Sukhbold16_CCSNe_expl, SN_exp_book, DR2}, which may determine the explosion energy and the explosive nucleosynthesis (cf., Sect.\,\ref{sec:expl}). 
The decay of iron-group elements, in particular of $^{56}$Ni, powers the bell-shaped region in the light curves for stripped-envelope SNe, as well as the radioactive tail at late times. As all our pre-explosion models are derived from the same initial mass, the core mass is not a parameter in our models.

We can obtain a good idea of the expected SN display based on existing SN models. 
For a fixed explosion energy and nickel mass, the key
parameters of hydrogen-rich models are the model radius and the envelope mass, where larger radii lead to larger peak (i.e., plateau) luminosities, and larger envelope masses to longer-lasting and dimmer
plateau phases \citep{Popov1993_IIP_Plateau_modeling,Young_2004_II_LC, Moriya2023_SyntheticRSG}.  

A sufficiently large amount of hydrogen is required for the SN to show hydrogen lines in its spectra, and larger radii will allow for bright light curves and an extended plateau phase. Finally, the envelope's composition will also affect the opacity and emissivity of the material, further affecting the appearance of the SN.

\subsection{Properties of the primary star}
By the time of collapse, binary mass transfer has brought about many changes in properties of the primary (\Tab{table:data}), most drastically in its envelope mass. Generally, the final envelope mass of the primaries remains larger for higher $P_\mathrm i$ and higher $\qi$ (\Fig{fig:Pq-plot}). The parameter space considered in this paper covers the entire range of envelope masses, from the non-RLOF models which never underwent any mass transfer, to the set of initially tighter models which lose their envelopes completely. 

The most extended pre-SN models occur in the initially tightest binaries which exclusively undergo \case C mass transfer (see \Fig{fig:Rad_Teff_vs_Menv_preCC}, and \Tab{table:data}), which retain envelopes of $M_\mathrm{env}\sim2.6\Msun$ and extend to $\sim 970\Rsun$. The reason why the largest progenitors have intermediate envelope masses stems from two competing effects. Initially wider systems allow the progenitor star to develop a larger envelope prior to mass transfer, but at the same time the more mass is lost during mass transfer the wider the primary's Roche lobe (and thus radius) can become. \EDIT{Overall, any of our primary models that retains an envelope  develops a radius of at least $200\Rsun$, since their envelope masses exceed $0.08\Msun$, with at least $0.01\Msun$ of hydrogen. To obtain compact, hydrogen-free SN progenitors from models which undergo mass transfer shortly before core-collapse (i.e., Case\,BC and Case\,C) appears to require a pre-SN common envelope evolution, while stellar wind mass-loss from the stripped donor star may produce this frequently in Case\,B systems (cf., Fig.\,\ref{fig:Pq-plot}).}

The approximately similar luminosity between these models results in a similar (albeit reversed) trend in terms of the progenitor effective temperature prior to collapse (\Fig{fig:Rad_Teff_vs_Menv_preCC}). The models with the largest radii are also the coolest, reaching surface temperatures as low as $3\,034$K (114K cooler than the non-interacting model), while the hottest ones are those with the smallest envelope masses, which are yellow super-giants.

Overall, effective temperature and radius correlate strongly with envelope mass (cf. \Fig{fig:Rad_Teff_vs_Menv_preCC}). The models with $0.15\Msun\simle M_\text{env}\simle 0.4\Msun$ add some scatter around the expected trend line, which, however, may be due to numerical noise in the models during the last $\sim 20\text{yr}$ prior to core collapse (see \Fig{fig:R_Teff_preCC_smooth} and \App{sec:App_primary}).

\subsection{Explosion energy and nickel mass}\label{sec:expl}
 Here, we obtain estimates for the explosion energy $E_\text{exp}$, the mass of nickel ejected $M(^{56}\text{Ni})$, the gravitational mass of the remnant $M_\text{ns}$, and its birth kick velocity $v_\text{kick}$. 
We apply the method of \citet{MHLC16} to our models with $\qi=0.95$ and different $P_\text i$, which is a representative sample of our entire model set. With the exception of the three models that crashed after core  silicon burning (cf., \Tab{table:data}), they all reached iron-core infall velocities of $1\,000\kms$. 

\EDIT{The explosion properties are sensitive to the CO-core mass \citep{Schneider_preSNevo_stripped_2021}. Our primary star models have the same initial mass and managed to retain a H-rich envelope (except for the pure Case\,B models). Therefore, the He-cores end up having very similar pre-SN masses (cf., Tab.\,\ref{table:data}), and also the final CO-core masses differ by less than 5\%. Consequently, our Case\,BC and Case\,C models can be expected to have a similar nickel mass, explosion energy and remnant mass, regardless of the envelope surrounding the core, and thus regardless of the mass-loss history of the system.
For this reason,} we quote here the average values across the different models. Using the parameters in the applied method as adopted in \citet{DR2}, we obtain $M(^{56}\text{Ni})=(4.02\pm 0.83)\times 10^{-2}\Msun$, $E_\text{exp} = (0.97\pm 0.15)\times 10^{51}\text{erg}$, $M_\text{ns} = (1.36\pm 0.03)\Msun$ and $v_\text{kick}=(313\pm 67)$\,km\,s$^{-1}$, where the scatter is within $20\%$ of the mean value. Changing the adopted input parameters by $5\%$ only slightly increases the spread, while leaving the average unaltered.
 
We define the ejecta mass of our models as $M_\text{ejecta}=(M_\text{He-core}+M_\text{env})-M_\text{ns}$. Given the high kick velocities, it is likely that the binary system will break up after the explosion of the primary.

\subsection{Expected SN types}\label{sec:Supernova_Types}

Based on \Fig{fig:Rad_Teff_vs_Menv_preCC}, we expect a one-dimensional family of SN light curves from our models, spanning all the way from \Type IIP-like light curves with extended plateaus for the higher hydrogen envelope masses ($\sim$\,7\,$\mso$), to \Type IIL SNe exhibiting fast-declining light curves (from a few down to
$\sim$\,0.3$\mso$) and \Type IIb SNe (with $M_\text{env}$ in the range $\sim 0.3\dots 0.001\mso$) to \Type Ib SNe for fully stripped models \citep{Morozova15_LCofCCSNe_SNEC, DessartHillier19_IIPLC, HillierDessart19_IIP_LC_spectra, Hiramatsu_IIP_partStripped, Dessart2024_widebinaryIIP_from_Ercolino2024}. \Figure{fig:Pq-plot}  highlights where in the considered parameter space to expect the different SN types from our models.

These quoted boundaries distinguish between different orders of magnitudes of the mass of hydrogen and are only meant qualitatively. The boundary between \Type Ib and IIb SNe has been discussed in the literature. \citet{Hachinger2012} argue that hydrogen lines disappear from the SN spectra with $M_\text{H}\simle0.033\Msun$. \citet{Luc2011} explores \Type IIb SN models that show strong H$\alpha$ lines even when the hydrogen mass is as low as $0.001\Msun$ and the same is found for the larger grid of models presented in \cite{Dessart2015_IIb_Ibc_RadTransf, Dessart2016_IIb_Ibc_RadTransf2}. Only five models in our grid fall between these definitions (cf. the yellow models in \Fig{fig:Pq-plot}), and using the boundary proposed by \citet{Hachinger2012} or that of \citet{Luc2011} would merely shrink the parameter space for \Type IIb in favor of \Type Ib SNe. However, the uncertainty in the implemented physics (cf. \EDIT{Appendix\,\ref{sec:APP_uncertanties}}) may result in the boundary between \Type IIb and \Type Ib SNe to be found at lower orbital periods than shown in our models.

\subsection*{\Type IIP/IIL SNe}
All our models with initially wide orbits retain enough hydrogen to be classified as either \type IIP or IIL SNe. More specifically, this includes the models not undergoing RLOF, all the systems undergoing stable \case C mass transfer, and those undergoing \case BC mass transfer with the highest $P_\text i$ and $\qi$. 

\Type IIP progenitors retain an envelope of at least $1\Msun$, which corresponds to an extension of more than roughly $900\Rsun$  and an effective temperature of less than $3.1\,$kK. (\Fig{fig:Rad_Teff_vs_Menv_preCC}).
\Type IIL progenitors from our set of models also all appear as RSG by the time of collapse, as their effective temperatures are found within $3.1-3.5\,$kK and their radii between $750-900\Rsun$. It follows that they would be difficult to discern from pre-explosion photometry alone, and only light-curve modeling would thus allow to distinguish them. 

The SN light curve and spectra of models that undergo \case C RLOF will inevitably be affected by the dense CSM surrounding them (cf., Sect.\ref{sec:SupernovaCSM}). 
As such, the only systems that explode as non-interacting \type IIP SNe will be the models which never undergo RLOF, as well as systems that merged during \case B RLOF. This means that within the progenitor initial mass range probed by our models, we do not expect any \Type IIL SNe without exhibiting interaction with the surrounding CSM.  

\subsection*{\Type IIb SNe}
In our models, \Type IIb progenitors can have a wide range of surface properties prior to collapse, which clearly scale with the envelope mass (cf. \Fig{fig:Rad_Teff_vs_Menv_preCC}; see \Appendix{sec:App_core_envelope} for how this is affected by the definition of the envelope mass). They can exhibit pre-explosion radii between $200$ and $700\Rsun$, from the least to the most massive envelopes, and it is likely that smaller radii might be achieved if we had a more refined grid of models. The effective temperatures also vary significantly, going from $6.0$kK for the models with the least massive envelopes to $3.5$kK for the most massive ones. 

\Figure{fig:Rad_Teff_vs_Menv_preCC} compares our models with observed \Type IIb pre-explosion stars (see \Tab{tab:SNe}). It shows that estimates of the progenitor radii (from both pre-explosion images or light-curve modeling) and envelope masses lie within the region traced by our models. 
The observed \type IIb progenitors are found in the region of the diagram which is mostly populated by our \Case B models, and by the shortest period \Case B+C models (\Fig{fig:Pq-plot}). This suggests that most of the observed SNe shown in \Fig{fig:Rad_Teff_vs_Menv_preCC} are not expected to undergo significant CSM interaction (see the right panel in \Fig{fig:Pq-plot_zoom}).

 The luminosities of our models with low envelope masses ($60.5-65.0\text k\Lsun$) are lower than of all of the progenitors reported in \Tab{tab:SNe}, but they are compatible within $1\sigma$ to those of \SN{2008ax} and \SN{2011dh}. Light-curve modeling of these two SNe indeed yielded that they may share similar progenitor masses ($4-5\Msun$ for \SN{2008ax}, \citealt{Folatelli_2008ax} and $3-4\Msun$ for \SN{2011dh}, \citealt{Bersten_2011dh}) to our models ($\simle 4\Msun$). Similar analysis on the progenitor masses of \SN{2016gkg} ($4-5\Msun$, \citealt{Bersten_2016gkg}) and \SN{1993J} ($3-6\Msun$, \citealt{1993J_Woosley}) yielded masses compatible to our progenitors, even though their pre-explosion luminosities are higher by a factor of two. Finally, estimates for \SN{2013df} place both the pre-explosion luminosity and final mass  ($5-6\Msun$, \citealt{Szalai2016_2013df}) to higher values than what our models predict.
While our models were not tailored to fit the mentioned \Type IIb progenitors, the dependence of their final radii and surface temperatures on their envelope mass appears qualitatively in good agreement with the observations.

\subsection*{\Type Ib}
All the \case B models that managed to shed their entire envelope via wind can be expected to end as \Type Ib SNe. The models computed here explode with $R\simeq 12\Rsun$ and $T_\text{eff}\simeq 25$kK. 

Helium stars with smaller mass experience a stronger expansion after core helium burning and could therefore produce cooler SN progenitors and even undergo \Case BC mass transfer \citep[e.g., ][]{Kleiser2018_Ibc_He_giants, Woosley2019_Hestars, WuFuller_IbcLMT}, which could be the case for models with initially lower $P_\text i$ and $\qi$ than our models exploding as \Type Ib SNe.

\subsection{Properties of the secondary star}
\label{sec:preSN_secondary_star}

The secondary star, being less massive than the primary star, evolves on longer timescales, and as such it remains in earlier stages of evolution, and the more so the lower its mass. Its evolution is followed until the point at which the primary star reaches core collapse (in some cases where numerical problems arise, its evolution is stopped during RLOF sometime after the primary depletes carbon, $\simle50$yr prior to the primary's collapse; see \App{sec:App_secondary} for details). 
For models with $\qi=0.95$, the secondary's evolution lags only slightly behind the primary's, and it reaches the TAMS $\sim 9\,000\years$ before the primary's collapse. The expansion following core hydrogen exhaustion brings the secondary to a radius of $\sim\,160\Rsun$, which would have been enough to also fill its Roche lobe if the orbit were initially tighter. This would have also occurred if the secondary's mass was even higher (but still smaller than the primary).
For lower initial mass ratios, the secondary is always still on the main-sequence by the time the primary reaches core collapse.  

Regardless of their evolutionary stage, our secondary stars only accrete a small amount of mass as it is enough to spin them to critical rotation. During the first mass transfer event (whether it is \case B or \case C) this amounts to less than  $0.25$\,$\Msun$ for all models, while during the second mass transfer event (if there is one) only less than $\sim$\,0.03$\Msun$ is accreted (cf. \Fig{fig:DeltaM_C}), as the star did not have enough time to spin down via wind mass-loss. 
Following the SN explosion of the primary star, the high rotation rate of the secondary star will be the main telltale sign of the system having undergone RLOF. 

We do not consider any influence of the secondary star on the visible manifestation of the SN explosion of the primary.

\section{Expected interacting supernovae}\label{sec:SupernovaCSM}

When the interaction between SN ejecta and CSM is considered, the classification proposed above may be significantly altered, in particular, if the interaction power dominates the various power sources (i.e., decay power and release of shock-deposited energy prior to shock breakout). A model identified as a \type IIP SN in the absence of CSM could turn into a \Type IIL with IIn signatures at early times. The complex CSM configurations produced in our grid of models probably leads to a wide range of light curve and spectral evolution, and similarly a diversity in SN classification (although most of our models would be of \Type II due to the systematic presence of hydrogen in the CSM). One further issue is the shortcomings of SN classification. For example, only the most extreme CSM configurations lead to a \Type IIn classification, while ejecta--CSM interaction is without doubt much more universal (see, e.g., \citealt{DessartHiller2022_IIP}).
Below, we review the various CSM configurations associated with our progenitor models at the time of core collapse and discuss the potential impact of such a CSM on the SN radiative properties.

\subsection{RSG radii and ``flash spectroscopy''}
\label{sec:RSG_radii_flash}

In our wide binary models, the primary stars are RSGs.
The structure of the atmospheres and the outflows from RSGs are not well understood. Here, we demonstrate the range of possibilities with some simple considerations. For this, we first consider an isolated RSG.

Using the wind mass-loss recipe described in Sect.\,\ref{sec:wind_method}, our $\sim 12.6\mso$ primaries have a steady-state wind mass-loss rate of $\sim 3\times10^{-6}\msoy$ (cf., \Fig{fig:MdotC_q0p95}).
With a radius of $\sim 1000\rso$, and typical densities at the outermost grid point of the MESA models (defined by a Rosseland optical depth of 2/3) of $\sim10^{-9}\gcc$, this leads to average outflow velocities at this point of the order of 3\,cm\,s$^{-1}$. The sound speed here is $\sim 10^5$ times larger. 
This implies that while the radius of the MESA models -- which is used to determine the mass transfer rates -- corresponds to the edge of the optically thick stellar body, it does not represent the edge of the hydrostatic part of the star. 

\begin{figure}
    \includegraphics[width=\columnwidth]{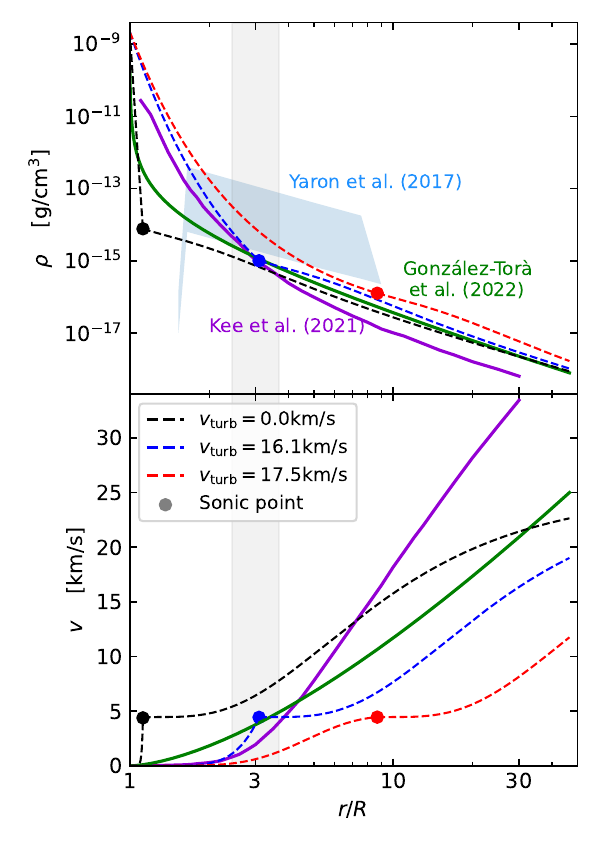}
    \caption{ \EDITL{Circumstellar density (top panel) and velocity (bottom panel) as a function of the distance from the outer boundary} of our $12.6\mso$ single star RSG model 1 year prior to collapse. Its
    radius as derived by MESA is $R=897\rso$, and
    its wind mass outflow rate $3.3\times 10^{-6}\msoy$. The three dashed lines correspond to an isothermal subsonic atmosphere which includes various turbulent pressure
    contributions (see legend), which transition to a $\beta$-wind law with $\beta=5$ and $v_\infty=25\text{km}/\text{s}$
    when the local velocity reaches the sound speed (marked by a thick dot). The purple line corresponds to the fiducial turbulent pressure-driven RSG wind model
    of \cite{Kee_Atmo_Pturb_2021}, while the green line represents the semi-empirical outflowing atmosphere model of \cite{Gonzalez-Tora_RSG} which fits the interferometric observations of
    the Galactic nearby RSG HD\,95687.
    The blue area represents the constraints for the circumstellar density distribution derived by \cite{Yaron_Flash} based on the flash-spectroscopy of
    the ordinary \Type IIP \SN{2013fs}.
    The vertical shading marks the typical orbital separation found in our binary models at core collapse. 
    }
    \label{fig:CSM}
\end{figure}

Therefore, an optically thin hydrostatic layer extends our fictitious star, and one can integrate the hydrostatic equation, assuming an isothermal structure, up to where the velocity reaches the sound speed to estimate its extent (cf., \Fig{fig:CSM}). The result is that the hydrostatic part of the star is extended by several pressure scale heights, which, based on the structure of our MESA model, amounts to $\sim 10$\% of the radius.
This extension is essentially accounted for in the mass transfer scheme adopted in our work (Sect.\,\ref{sec:RLOF_method}).

On the other hand, the outer envelopes and atmospheres of RSGs are strongly affected by the turbulence imposed by convection. 
It has been argued that this turbulence provides a major contribution to the pressure in these layers \citep{Jian_Huang_Pturb_II, Stothers_Turbolence, Grassitelli_Turbolence,Goldberg2022_3Dhydro_turb} and that it may help or even cause the driving of the RSG wind \citep{Jian_Huang_Pturb_I, Jian_Huang_Pturb_II, Jian_Huang_Pturb_III, Kee_Atmo_Pturb_2021}. 
For a constant turbulent velocity (\citealt{Kee_Atmo_Pturb_2021} propose $15\kms$), and assuming isotropic turbulent pressure as $P_{\rm turb}\simeq {1\over 3} \rho v_{\rm turb}^2$,  the hydrostatic equation is again easily integrated. \Figure{fig:CSM} shows that, in this case, the sound speed is reached only at a large distance from the edge of the optically thick stellar surface. This is consistent with the wind models of \citet{Kee_Atmo_Pturb_2021}, where this happens in their fiducial case ($10\mso$) at five times the stellar radius. If turbulent pressure is taken into account, the subsonic optically thin region may contain as much as $0.2\Msun$ of material.

Also, interferometric observations of RSGs imply extended quasi-stationary envelopes \citep{ArroyoTorres_whylargeRSG}. \citet{Gonzalez-Tora_RSG} have analyzed the nearby $\sim\,15\mso$ RSG HD\,95687, for which they derived the outflow structure shown in \Fig{fig:CSM}. While in their solution, the sonic point is reached at three times the RSG model radius, it is not free of ambiguities, as a velocity and temperature structure needs to be assumed in order to interpret the interferometric visibility functions. 

Such large extensions of RSGs can have important effects in two main areas. Firstly, if the quasi-hydrostatic size of RSGs was five times larger than the stellar structure radius provided by MESA, mass transfer in a given binary model would start much earlier, while, on the other hand, the upper initial period limit for interaction in circular binaries would shift from $\sim 20\,\text{yr}$ to $\sim 75\,\text{yr}$ for our selected primary star mass -- where eccentric systems could even interact at much longer periods. 
In addition, especially when the wind speed is smaller than the orbital speed of the mass gainer, \citet{Modamed_Podsiadlowski_WindAccretion_2007, Modamed_WindAccretion_2010} showed that the companion stars can capture much more mass from the primary's wind than what is expected from Bondi-Hoyle accretion \citep[cf.,][]{Saladino_2018_WindAccretion}. 

Secondly, quasi-hydrostatic, optically thin extensions of RSGs can strongly affect the first days of the display of the SN. For our most extreme model in \Fig{fig:CSM}, the sonic point at $\sim\,8\,000\rso$ implies a light travel time of $5\,\text{hours}$, and it will be reached by the SN shock after $\sim\,3\days$ (assuming a shock velocity of $20\,000\kms$).

In \Fig{fig:CSM}, we compare the theoretical and semi-empirical density profiles with that deduced from the
very early time spectra (``flash-spectroscopy'') of the normal \Type IIP \SN{2013fs} \cite{Yaron_Flash}. 
Noticeably, all but perhaps the isothermal model with the largest assumed turbulence velocity fall somewhat short in reproducing the density structure surrounding \SN{2013fs}. However, \citet{Kee_Atmo_Pturb_2021} and \citet{Gonzalez-Tora_RSG} are considering average, i.e., core-helium burning, RSGs, while at the time of core collapse, RSGs have an elevated $L/M$-ratio, and thus likely a denser wind than during core helium burning  (cf., \App{sec:App_uncertanties_winds}). 
Similarly, the adopted RSG mass-loss rate in our MESA model is empirically calibrated to observed RSGs, and the adopted rate of $\dot M=3.3\times 10^{-6}\msoy$ 
may also underestimate the true pre-SN mass-loss rate for a $\sim 12\mso$ RSG. We argue in \App{sec:App_uncertanties_winds} that the pre-SN mass-loss rate may be up to ten times larger than what is assumed for all the curves in \Fig{fig:CSM},
which would shift them upwards by one order of magnitude, well into the regime of agreement with \SN{2013fs}.

It has been concluded from very early time spectra of several \Type IIP SNe that many, if not most RSGs, are enshrouded in a rather dense envelope at the time of core collapse \citep[see, e.g., ][]{Bruch2022_CSM_II}. \citet{Yaron_Flash} concluded for \SN{2013fs} that this envelope is produced by a $100\kms$ outflow with $\dot M\simeq 10^{-3}\msoy$ over the last year in the life of the pre-SN star. Our considerations above may allow for the less spectacular interpretation that rather ordinary RSG winds, with $\dot M\simeq 10^{-5}\msoy$, can perhaps accelerate slowly enough to reproduce the results from flash spectroscopy \citep[see also discussion in][]{Dessart2017_RSGexpl,Moriya2017_CSM_wind_SNLC}. In this case, the flash spectroscopy results would strictly speaking not indicate CSM interaction, as the observed narrow lines would be produced by a hydrostatic and quasi-stationary (even though turbulent and possibly pulsating) part of the progenitor star.

\subsection{Supernovae during stable mass transfer}\label{sec:interaction_with_RLOF_CSM}

In many of the analyzed models, mass transfer is continuing up to the core collapse of the RSG primaries (cf. Figs.\,\ref{fig:MdotC_q0p95} and \ref{fig:MdotC_dq_1413_1995}). 
As the mass gainer in our binary models is quickly spun up, nearly all the transferred matter is lost from the binary system in our models (cf., Sect.\,\ref{sec:RLOF_method}).
While the mass transfer efficiency, and thereby also the rate of mass ejected by the mass transfer process into the CSM, is uncertain \citep{Claeys_b, Langer_review_2012, Sravan_b}, the order of magnitude of the latter is in any case expected to be the same as that of the mass transfer rate \EDIT{(cf., Sect.\ref{sec:RLOF_method})}. 
From this effect alone, we expect the CSM density in systems with ongoing mass transfer to be much higher than, e.g., in single RSGs. 

The range of mass transfer rates spans all the way from values near those of standard RSG wind mass-loss
rates (BC794-0.95; \Fig{fig:DeltaM_C}) to more than 0.1$\msoy$ (C1995-0.60; \Fig{fig:MdotC_q0p95}), where models
with even larger mass transfer rates are expected to undergo a common envelope phase (cf., Sect.\,\ref{sec:unstable_caseC}). The peak in the mass transfer rates for systems that avoid a common envelope occurs typically $\sim\,20\kyr$ before the SN, with rates between $10^{-5}\dots 10^{-3}\msoy$ at the time of the SN explosion (see \Tab{table:data}). 

The total mass lost during this thermal timescale pre-SN mass transfer ranges from to a few tenths to seven solar masses (cf., \Fig{fig:Pq-plot_zoom}, first panel). 
The larger values occur in the pure \Case C systems,
since the partially stripped \Case B systems that undergo a final \Case C mass transfer have lost much of their envelope already during \Case B mass transfer about $10^6\years$ earlier. 

The fate of the ejected matter is uncertain. \EDIT{Hydrodynamic simulations of the outflowing matter would be required, but do not exist for our situation, and they would need to cover several thousand orbital periods, which is difficult on principle grounds. However, we can derive an expectation from two types of models in the literature.} 

\EDIT{Roche lobe overflow or wind mass transfer from a low-mass red giant to a compact companion star in a binary system has been modeled in several works  \citep[e.g., ][]{TheunsJorissen1993_WindAccretionBinaries,MastrodemosMorris1998_BipolarNebulaeBinaries, Walder2008_RSOph_Nova_accretion, BoothMohamedPods_2016_innertorque_CSM_RSoph_Ia}. These simulations show that the inefficient binary mass transfer leads to a mass outflow from the binary in which most of the mass is confined towards the orbital plane. In these simulations, hydrodynamic and tidal effects accelerate the gas to velocities beyond the escape speed, such that the mass density near the binary system remains relatively low. However, in these simulations, the wind particles are injected with velocities exceeding the orbital velocities of the stars (typically $20\kms$), and the lost mass is negligible in the mass budget of the system.}

\EDIT{In our models, the situation is different.}
As the amount of lost mass is substantial, this material can be expected to have a significantly smaller outflow velocity than the ordinary RSG wind. \EDIT{Furthermore, the outflows of red supergiants may be very slow out to significant distance (e.g., comparable to the orbital separation; cf., Fig.\,\ref{fig:CSM}), and thus may remain much slower than the orbital motion. In this respect, it is useful to consider the circumbinary disk models of \citet{ArtymowiczLubow1994_Dynamics_Binary_Disk_Interaction}, who modeled the effect of a central binary system on an otherwise steady, Keplerian disk. They showed that the central binary creates an inner cavity or gap by pushing most of the closest particles to a bigger distance, and that most of the disk mass accumulates in a circumbinary torus at the outer edge of this gap, at a radius of $2\dots3$ orbital radii (see, e.g., their Fig.\,9).}

\EDIT{The situation in our binaries could be comparable, when considering that matter is released from the Roche lobe of the red supergiant in near Keplearian motion and spreading in the orbital plane. } It may form a thick circumbinary torus, with the bulk of the matter remaining close to the binary system, with perhaps some matter falling back, some being driven further out. With a pre-SN stellar radius of $\sim 6 \times 10^{13}\,$cm and an orbital separation of $\sim$\,10$^{14}\,$cm, the inner radius of a circumbinary torus should be beyond $\sim 3 \times 10^{14}\,$cm, \EDIT{according to \citet{ArtymowiczLubow1994_Dynamics_Binary_Disk_Interaction}. A stationary situation may not be achieved in reality, as internal viscosity would tend to spread the torus, with some mass moving out, some back to the binary. However the timescale for that, when using our
binary geometry and the standard viscosity parameter of $\alpha=0.1$,
equals to several thousand orbital periods \citep{Pringle1981_AccretionDisks, Martin2023_Mergers_CircumbinaryDisks}}

\EDIT{Circumbinary tori are a widespread phenomenon, found or expected in binaries that are young \citep[e.g.,][]{Price2018_CircumbinaryDisks, Duffell2020_AccretionandTorque_in_CircumbinaryDisks}, evolved 
\citep[e.g.,][]{KashiSoker2011_EvolvedCircumbinaryDisk, Kluska2022_Disks_around_evolved_Stars}, or even supermassive \citep[e.g.,][]{Franchini2021_CircumbinaryDisk_SupermassiveBHMerger}. A potential observed case which may relate
to our binary models is the spatially resolved dust torus of the red supergiant \object{WOH G64} in the LMC \citep{Ohnaka2008_resolved_torus_around_RSG}. 
With an inner torus radius of $\sim 2\times10^{15}\,$cm, it might support the scenario above.}

\EDIT{In view of this situation, which leaves plenty of uncertainty,}
we assume in the following that the bulk of the material lost from the binary during the pre-SN mass transfer is at $r \simeq 10^{15}\,$cm, which means that SN ejecta, flying at $10\,000\kms$, will reach it within $\sim$\,12 days.
As mass transfer is ongoing, the interaction of the SN with CSM may start earlier. \EDIT{Notably, \cite{Smith_McCray_2007_SN2006gy_CSM_10e15cm} in their analysis of the Type\,IIn
SN\,2006gy, which produced a radiated energy of $\sim 10^{51}\,$erg, found
consistency with the observed supernova properties by adopting a radius for a massive circumstellar shell of $10^{15}\,$cm.}

 We can estimate the total energy that the ejecta loses in the form of radiation as it rams into the CSM. Following the work of \cite{DRAD18_GRB_SLSN_Ic}, we assume conservation of momentum and an inelastic collision, resulting in an amount of kinetic energy being lost equal to \begin{equation}\label{eq:E_interaction}
    \Delta E = E_\text{kin,ej}f_Mf_v^2,
\end{equation} where $f_M=\frac{M_\text{CSM}}{M_\text{ej}+M_\text{CSM}}$ and $f_v=\frac{v_\text{ej}-v_\text{CSM}}{v_\text{ej}}$. Assuming that all of this energy is released as radiation and that the CSM is slow-moving compared to the SN ejecta, the fraction of kinetic energy that is converted in radiation is given just by the factor $f_M$, which is more significant for those systems where the CSM is more massive. Given an explosion energy of $\sim$\,10$^{51}$\,erg (see Sect.\,\ref{sec:expl}), and a cumulative luminosity from the light-curve on the order of $\sim 3 \times 10^{49}$\,erg \citep{Dessart2024_widebinaryIIP_from_Ercolino2024}, it is clear that the models where $f_M\simgr 0.01$ may exhibit significant effects on the light-curve. However, the actual structure and extent of the CSM will determine when and how strongly this effect will be visible in the light-curve.

 When distributing the mass that is lost during the final mass transfer phase evenly within 
$10^{15}\,$cm, we can obtain a density of $\rho= {3\over 4\pi} {\Delta M_\text{RLOF,C} 
\over r^3}$. 
When ionized, this may give rise to an optical depth of 
$\tau \simeq \kappa \rho r = {3\over {20}\pi} {\Delta M_\text{RLOF,C} \over r^2}$,
using $\kappa = 0.2\,$cm$^2$g$^{-1}$.
Below, we distinguish several situations based on these numbers.

\subsection*{\Case BC mass transfer}
In our \Case BC models, the \Case C mass transfer removes between 0.01 and 2\,$\mso$ during the last $10^4\years$ before the SN,
with a typical value of $\sim$\,0.2$\mso$. With the assumptions above, this would result in a CSM density of $10^{-13}\gcc$ and an optical depth of 20.
The average envelope mass of $0.5\mso$ leads to a typical mass of the SN ejecta of $3\mso$. Using Eq.\ref{eq:E_interaction} implies that in this situation, the SN-CSM interaction can transform less than 10\,\% of the SN kinetic energy into radiation. These models would therefore not qualify for Superluminous SNe.

On the other hand, the interaction power could still exceed the luminosity of the SN if it was exploding into a vacuum. 
However, this is certainly not expected for some of our models, for which the computed CSM mass is even a factor of 10 smaller (BC794-0.95, BC891-0.95, and BC1413-0.70). 

Our \Case BC models could correspond to the class of \Type IIn SNe which are thought to have a density configuration in which
the dense CSM extends out to $~10^{15}\,$cm, often assumed to be produced by wind with mass-loss rates of order $0.01\msoy$ during the last decades of the life of the SN progenitor. In these SNe, the IIn signatures are present for about one week and then disappear. The prototype for this is \SN{1998S} \citep{Leonard_1998S, Dessart2016_IIn}, and more recent
examples have been observed: \SN{2020tlf} \citep[an event with recorded pre-SN activity,][]{Jacobson_Galan_2022_precurso2020tlf}, and the already famous \SN{2023ixf} \citep{Jacobson_Galan_SN2023ixf, Jencson_SN2023ixf, Smith_SN2023ixf}.
We also suggest the \Type IIb \SN{1993J} to be in this category. It might fit to our \Model{B}{794}{0.95}, due to its small remaining hydrogen envelope mass ($0.28\mso$) and small expected CSM mass ($0.01\mso$). 

\subsection*{Stable \Case C mass transfer}

For our \Case C models, the predicted pre-SN CSM mass is one order of magnitude larger than for our \Case BC models. Here, we first focus on the models which avoid a common envelope phase, specifically on models with an initial mass ratio above 0.6. These models expel $1\dots 4\mso$ during their final evolution, which would result in a CSM density of $6\times10^{-13}\gcc$ and an optical depth of 200.
In some of the considered models (e.g., B2512-0.95), the mass of the SN ejecta is larger than that of the CSM mass; however, in some cases, they are rather comparable, such that perhaps as much as 50\% of the kinetic energy of the SN can be absorbed and radiated. 

As such the total photon output of the SN may exceed 10$^{50}$\,erg due to the interaction of several solar masses of hydrogen-rich ejecta with a CSM whose mass is larger or at least comparable to the ejecta mass. Such events are expected to be super-luminous (10 times more radiation emitted than in non-interacting SNe IIP), exhibit emission lines with narrow cores and electron-scattering broadened wings arising from the unshocked CSM and photo-ionized by radiation arising from shock. Examples are \SN{2010jl} \citep{Zhang_SN2010jl} or \SN{2017hcc} \citep[e.g., ][]{Moran_2023}. Modeling of the light curve and spectra of \SN{2010jl} suggested a CSM configuration formed through a mass-loss rate of $0.1\msoy$ and a velocity of 100\,$\kms$ for $\sim 30\,$yr \citep{Fransson_2010jl, Dessart_IIn_2015}.

\subsection{Supernovae in a common envelope}

Our lower mass-ratio \Case C models develop such high mass transfer rates that a common envelope phase is expected. However, all these models do transfer $1\dots 2\mso$ before a common evelope forms, which happens in the most extreme case just $2000\,$yr before core collapse (\Fig{fig:MdotC_dq_1413_1995}). This will lead to a high CSM density similar to the cases discussed above.  

For the description of the common envelope evolution, we have to rely on analytic estimates (see Sect.\,\ref{sec:unstable_RLOF}). While these are uncertain, the prediction that the lowest mass-ratio systems will merge, and that the ones with the highest mass ratio have the energy to eject their common envelope (\Fig{fig:Pq-plot}) appears reasonable. In our calculation, the critical initial mass ratio 
for merging is about 0.2. On the other hand, the transition between systems that merge and those where the orbit decay produces enough energy to eject the envelope is perhaps smooth. 

The reason is that the duration of the common envelope phase may be similar to the remaining lifetime of the RSG primary at its onset. 
Our estimate of the duration of the common envelope phase (Eq.\,\ref{eq:tCE} in Sect.\,\ref{sec:unstable_RLOF}) leads to the prediction that in some of our models, it is still ongoing at the time of core collapse,
while in others it is already finished. We therefore discuss both cases here.

For the case where the common envelope phase is still ongoing when core collapse occurs, we have no robust estimate of the amount of CSM \EDIT{that} is already ejected at that time, except for the $1\dots 2\mso$ of matter which have been transferred before the common envelope phase started. 
\EDIT{In particular considering systems which encounter strong fall back (cf., Sect.\,\ref{sec:unstable_RLOF}) leaves a large uncertainty on the amount of envelope mass which is lost before the supernova occurs.}
We reflect the corresponding uncertainty for the SN-CSM interaction 
by considering the whole possible pre-SN envelope mass range, i.e.,
$\sim 0.2\dots 7\mso$, for the corresponding models in Figs.\,\ref{fig:Pq-plot} and\,\ref{fig:Pq-plot_zoom} and \Tab{table:data}.

However, the time from the onset of the common envelope phase to the SN explosion is clearly long enough to severely disturb the hydrostatic and thermal equilibrium of the SN progenitor. This implies that the progenitor, as well as the mass outflow rate from it, may be highly time-variable. We suggest that this variability could be related to the pre-SN activity observed in some interacting SNe \citep{Strotjohann2021_interactionpoweredSN}. It might apply to some SNe\,IIP with early time signatures of interaction (e.g., \SN{2020tlf}; \citealt{Jacobson_Galan_2022_precurso2020tlf}). The situation may be similar in binaries in which there is not enough energy to eject the common envelope (e.g., C1778-0.20, C1585-0.15, C1413-0.10). Also here, $\sim 1\mso$ of matter may be spilled before the onset of the common envelope phase, and the SN progenitor will
be far from thermal equilibrium at the time of core collapse. In this situation, the derivation of progenitor properties from pre-SN photometry, including the progenitor luminosity and mass,  could give spurious results.

On the other hand, when the common envelope phase ends with an envelope ejection before the SN event, it will do so very shortly before, such that the bulk of the envelope mass -- here $\sim 7\mso$ or so -- is still close to the binary system. As it dominates the mass budget, with only about 
$2.5\mso$ of SN ejecta, this may form a classical case of a superluminous SN, with 80\% of the kinetic energy transformed into radiation. An example model system may be BC1788-0.55. 

Perhaps, broad lines are never seen in such supernovae.  In the single-star scenario, this configuration was proposed to arise
from nuclear flashes in lower-mass massive stars, although in the crresponding models these flashes seem to occur too late, in association with a Si flash \citep[see][]{Woosley_Heger_2015_SiFlash}. 
The prototypes for this are \SN{1994W} 
\citep{Dessart2016_IIn}, and \SN{2011ht} \citep{Mauerhan_2011ht, Chugai_2011ht, Dessart2016_IIn}. 
In this category, we also have \SN{2006gy}, proposed as a Ia-CSM \citep{Jerkstrand_IaCSM}. Earlier interpretations of \SN{2006gy}
as pair-instability SN \citep{Woosley2007_2006gy} or Eta-Car analog (\citealt{Smith07_2006gy,Smith10_2006gy}) appear to have numerous points of tension.

Finally, if the time between the end of the common envelope phase and the SN exceeds the thermal timescale of the remaining stripped-envelope star, the direct SN progenitor may also be a blue supergiant.
This may offer a connection to objects that stand apart from all the cases discussed above. The prototype for this type is \SN{2009ip} \citep{Pastorello_SN2009ip, Fraser_SN2009ip,Margutti_SN2009ip, Smith_SN2009ip,  Fraser15_SN2009ip}. \citet{Smith_SN2009ip} and others again argue for a massive LBV progenitor, but there is no conclusive evidence for a very massive progenitor (this is usually inferred from the helium core luminosity of a star in hydrostatic equilibrium).

\section{Expected number of interacting \Type II SNe} \label{sec:SN_rates}

In the strict sense of the word, every \Type II SN interacts with the CSM, because all RSGs have a relatively dense 
wind. As explored in Sect.\,\ref{sec:RSG_radii_flash}, many, perhaps all, RSGs should also have extended stationary envelopes, which could produce interaction features, in particular narrow spectral lines, during the first day or two after core collapse. For our rate estimate below, we do not consider these cases but only include \Case BC and \Case C models corresponding to the colored pixels in \Fig{fig:Pq-plot}.

We perform a simple estimate of the number of interacting SNe normalized to the total core collapse SN rate, based on the parameter space we explore. 
We only consider binary systems with our chosen single star mass and adopt an initial binary fraction of one.

\Figure{fig:Pq-plot} predicts an interacting SN in 88 of 1224 binaries undergoing RLOF. If the initial parameter distribution was linear in $\log P_\text i$ and in $\qi$, and with about half of the other 1136 systems merging and giving rise to only one SN\,IIP per system, the other half producing one SN from a stripped star and one SN\,IIP, we would obtain 65\% \Type IIP, 30\% \Type Ib/c and 5\% interacting SNe. Of the latter, roughly half could have the SN light dominated by the interaction, including cases producing superluminous \Type IIn SNe. 

Here, in particular, the initial binary fraction and the merger fraction introduce uncertainty factors on the order of two. 
Moreover, the fraction of interacting SNe could be much larger than what we predict, due to three neglected effects which are discussed in the remainder of this section.

\subsection{Red supergiant extended envelopes}
\label{sec:extended_RSG_envelope}

We already include the Kolb scheme for RLOF, but this only accounts for a radius increase of 10\% (cf. black model in \Fig{fig:CSM}). If the hydrostatic radius of RSG was larger by a factor of five, then binaries with an initial orbital period of up to $\sim 75\text{yr}$ may interact. In the best case scenario, as in all of the binaries between what we have simulated and this new upper boundary undergo \case C RLOF, the number of interacting SNe can increase by $\sim 10-15\%$ at the expanse of the normal \Type IIP. This alone would push our estimates closer to the observed number in \cite{Smartt_rev_2009}. 

\subsection{Red supergiant pulsations}
\label{sec:RSG_pulsation}

Based on the recent Gaia data, \cite{MA23_StellarVariability_GaiaDR3} show that all galactic and Magellanic Cloud RSGs are semi-regular brightness variables. 
This implies that all RSGs are pulsating to some extent. 
Considering the best-studied examples, namely Betelgeuse and Antares, the typical variability timescale is several hundred days \citep[see e.g., ][]{Stothers1971_variableRSGs, Smith1989_Antares, Percy2014_Antares,Jadlovsky2023_Betelgeuse}, and spectroscopic velocity measurements identify velocity amplitudes on the order of 5\,$\kms$ \citep{Dupree2022_BetelgeuseDimming}. This implies radius fluctuations of ordinary RSGs on the order of 10\%.

These ordinary semi-regular pulsations are possibly triggered by the large-scale convection in the outer RSG envelopes \citep{Chiavassa10_Betelgeuse_RHD, Ahmad23_selfexcited_pulsations_AGB_RSG}. 
As showcased by Betelgeuse's Great Dimming \citep{Montarges21_Betelgeuse_Great_Dimming}, the stochastic behavior of these pulsations may also occasionally enhance the RSG mass-loss rate by factors of ten or more
\citep{Dupree2022_BetelgeuseDimming}. 

Furthermore, RSG pulsations have been found in
hydrodynamic stellar evolution models \citep{Pulsation, Yoon_Cantiello_2010, Moriya_Langer_2015_Pulsations}. Due to the 1D nature of these calculations, only radial oscillations have been identified; however, the excitation of non-radial pulsations can not be excluded. These single-star models consistently identify the $L/M$-ratio as a driving factor, in the sense that faster growth rates are found for higher $L/M$-values. 
\cite{Pulsation} confirmed the pulsational instability,
pulsational periods and growth time scales found in the
non-linear pulsation models with linear stability analyses, which excludes numerical effects as the cause of the pulsations in the numerical hydrodynamic models.

The non-linear models found pulsation amplitudes of up to 50\% in radius, at which time the numerical calculations broke down. \cite{Moriya_Langer_2015_Pulsations}, who investigated models of initially metal-free very massive RSGs, explored the damping of the pulsation by enhanced mass outflows, and found a saturation of the growth of the pulsations at radius amplitudes of roughly 50\%, for mass-loss rates on the order of $10^{-4}\dots 10^{-2}\msoy$. 

The models presented here were computed without hydrodynamics, such that their pulsational properties are not examined. 
However, as many of them experience a significant stripping of their hydrogen-rich envelope during \Case B and/or \Case C mass transfer, their mass is strongly reduced while their luminosity remains largely unaffected (cf. Table \ref{table:data}).
Consequently, their $L/M$-ratio is boosted to the level of $\sim 16\, {\rm k}\lso/\mso$, or analogously their value of $\mathcal L={T_\mathrm{eff}^4}/{g}$ is increased by a factor of two or more, which puts them into the regime of the most violently pulsating models of \cite{Pulsation}, with growth timescales as short as 40\% of their dynamical timescale.

Whereas in single stars, pulsations may enhance the density of the outflow and thus the RSG mass-loss rate, the consequences in a binary system may be more drastic. Most of our pre-collapse models are either close to filling their Roche lobe or undergoing RLOF and mass transfer. 
In these models, a fluctuating RSG radius would obviously increase the pre-SN RSG mass-loss drastically.

A second effect of the large amplitude pulsations expected for the partially stripped RSGs is that they would lead to temporary RLOF in binary systems which would be too wide to induce mass transfer otherwise. For example, our initial orbital period limit for interaction in circular binaries with $12.6\mso$ primaries of
$\sim\,2800\days$ would increase to $\sim\,7900\days$ ($\sim\,22\years$) if the maximum radius of the pulsating RSG would be twice that in quiescence. 

The Roche lobe filling and possible mass transfer by a (semi-regular) pulsating star is intermittent, and thus one might expect a smaller time-averaged mass-transfer rate in this case, compared to that of a non-pulsating Roche lobe filling RSG. However, also in this situation, the induced mass-loss from the RSG will lead to an increase in its (time-averaged)
radius, such that the loss of a substantial fraction of the envelope mass can be expected. These effects together are thus likely to expand the part of the initial binary parameter space in which RSG mass primaries are partially stripped.

\subsection{{Orbital eccentricity}}
\label{sec:initial_eccentricity}

Our models assume initially circular orbits. However, while in tight binaries tidal forces can quickly lead to circular orbits during the main-sequence evolution, systems with initial orbital periods in the range discussed here are likely to be eccentric at the time when mass transfer begins.
In this case, mass stripping from the more evolved primary star will only occur during the periastron passage. 
 
While solid predictions for this situation are not available, some insights may be derived from considering observed eccentric eclipsing RSG binaries, of which the well-studied system VV\,Cep is a prominent example. It consists of a RSG and an early B\,type main-sequence companion \citep{HagenBauer2008_VVCep_RSG}. With an orbital period of  $7430\days$ ($20.4\years$) and an eccentricity of 0.38,  the supergiant's Roche radius shrinks to less than $1\,000\rso$ near periastron, which triggers episodic mass transfer \citep{PollmannBennet2020_VVCep_spec}.

While VV\,Cep has only been studied for about half a century, its existence argues in favor of the long-term stability of eccentric RSG binaries with episodic mass transfer.  Since the RSG is only
stripped during a fraction of its orbital period, 
the time-averaged mass transfer rate is expected to be smaller than for continuous mass transfer, increasing the likelihood of partial stripping of the supergiant. Together with the fact that such partial stripping may then occur for initial orbital periods which are much larger than the upper period limit for interaction in circular orbits ($\sim\,2500\days$ for our models with $\sim\,12.6\mso$ primaries), may imply that many more RSGs end up just partially stripped than what is implied by our analysis of circular systems (i.e., \Fig{fig:Pq-plot}). 
Furthermore, the episodic non-conservative mass transfer may contribute  additional CSM structures to the already rich spectrum of possibilities, as well as lead to pre-SN activities of the progenitor star itself.

Together, the effects of larger RSG extensions (Sect.\,\ref{sec:extended_RSG_envelope}), RSG pulsations (Sect.\,\ref{sec:RSG_pulsation}) and finite eccentricities, could significantly increase the fraction of interacting SNe produced in wide binaries relative to the estimate provided at the beginning of this section.

\section{Conclusions}\label{sec:conclusions}
We investigated a set of more than 100 detailed long-period binary evolution models with MESA using a fixed initial primary star mass of $\sim$\,12.6\,$\mso$. 
These models initiate a strong binary mass interaction {from $30\kyr$ to $1\kyr$} before the primary star undergoes core collapse, either via \Case C or \Case B+C mass transfer, which is still ongoing at the time of the SN explosion.

We find that the widest systems undergoing RLOF manage to undergo stable mass transfer, even if only within a limited initial mass-ratio range, and sometimes only for part of the mass-transfer phase. The mass donors do not always lose the entire hydrogen-rich envelope in the process, but exhibit a continuous spectrum of final envelope masses, from completely stripped models in the tighter binaries to models containing progressively more massive H-rich envelopes in wider systems. If no CSM interaction were to occur, these models would cover the whole sequence from \Type Ib, IIb, IIL, to IIP SNe.

However, we find that the SNe expected from our model grid could have a significant, and in part dominant, power contribution from an ejecta--CSM interaction. In those models where the SN explosion
occurs during RLOF, the primary typically sheds a few tenths of a solar mass in the \Case BC systems, and between $2$ and $7\mso$ in our \Case C models.

The recent findings of \cite{Ouchi2017_IIb_RSG_progenitors} and \cite{Matsuoka_Sawada_BinaryInteraction_IIP_Progenitors} support our results for wide binaries undergoing stable mass transfer.
Although the shape and size of the CSM is not well constrained by our work, we expect a strong interaction with the SN ejecta, leading to a variety of short- and long-lived \Type IIn SNe in events that would otherwise (i.e., in the absence of CSM) have appeared as \Type IIP, IIL, or even IIb SNe.

We find that the final mass transfer in our \Case C models with companions of less than half the initial primary mass becomes unstable after several solar masses of material have been transferred.
We assume that this leads to the formation of a CE, for which we estimate the outcome
and timescale based on analytical recipes.  While these recipes are uncertain, the results imply an interesting range of possible outcomes.

In most of our binary models undergoing a phase of CE, the phase of CE evolution is expected to be in an advanced stage at the moment of collapse but not yet complete.
In some of those models, the orbital energy of the in-spiraling companion is found to be sufficient
to unbind the RSG envelope, while for the lowest initial mass ratios ($\simle 0.25$) it is not.
In most cases, neither the stage of CE ejection nor that of a merger is achieved before the core collapse of the RSG. Instead, the CE evolution is still ongoing during core collapse,
such that it may produce significant pre-SN activity. 

The CSM structure is difficult to estimate, but hydrodynamic CE models show that the
common envelope may not be lost in one episode, but rather in several episodes, with some matter also falling back. Between 1 and $7\mso$ of material is expected in the CSM, which would be conducive to a strong ejecta--CSM interaction.
In those of our models for which a CE ejection is estimated to occur before the SN, the massive CSM is expected to be found close to the SN, leading to a superluminous event. 

While our models leave many open questions, in particular regarding the CSM structure, they \EDITL{may} reveal just a small amount of the potential of wide binaries to contribute to interacting SNe. In particular, we identified three physical mechanisms, all of which had to be neglected in this paper. First, the part of the RSG envelope with highly subsonic average ourflow velocities may extend far
beyond the edge of the star as predicted by stellar models. If so, much wider binaries than investigated here could undergo late mass transfer and RLOF. 
Second, RSGs may pulsate with significant amplitudes, especially during their final evolution. This could again widen the parameter space for interaction, and lead to more partially stripped RSGs and
further enhanced CSM densities. 
Third, most of the wide binaries must be expected to be eccentric,
such that mass transfer could occur only episodically near periastron.  

These effects may imply that
the rate of interaction-dominated SNe exceeds our simple estimate of $\sim 5$\% of all core-collapse SNe. Furthermore, as these effects may be equally important in binary systems consisting of RSGs and compact companions, interacting SNe may provide important lessons for understanding the CE evolution channel towards tight double neutron star and black hole binaries \citep[see, e.g.,][]{Moreno2022_CEE_to_GW}. 

Finally, the processes ruling the hydrogen-rich interacting SNe may occur analogously in 
stripped-star binaries.
In corresponding models, when the hydrogen- and/or helium-deficient star is of sufficiently low mass \citep[$\simle 3.5\mso$;][]{Wellstein_2001_ContactMassive, Yoon2010_Ibc, WuFuller_IbcLMT}, it may expand enough to initiate mass transfer shortly before collapse. 

\begin{acknowledgements}

\EDITL{We thank the anonymous referee, whose comments and criticism helped to
improve important parts of this manuscript.} We also thank Alex de Koter, Markus Wittkowski, and Gemma Gonz\'alez-Tor\`a for the insightful discussions. This work was fostered by the participation of NL and LD at the MIAPP Workshop on {\em Interacting Supernovae} in February 2023.  \EDIT{This research was supported by the Munich Institute for Astro-, Particle and
BioPhysics (MIAPbP) which is funded by the Deutsche Forschungsgemeinschaft
(DFG, German Research Foundation) under Germany's Excellence Strategy –
EXC-2094 – 390783311. AE acknowledges the support from the DFG through grant LA 587/22-1.}
\end{acknowledgements}

\bibliographystyle{aa}
\bibliography{Reference_list}

\begin{appendix} 

\FloatBarrier
\section{Advanced evolutionary stages and numerical instabilities}\label{sec:App_CC}

\begin{figure}
    \includegraphics[width=\linewidth]{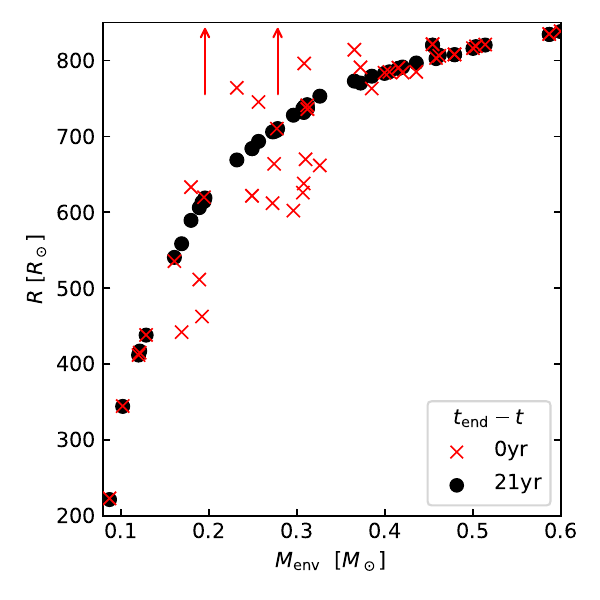}
    \includegraphics[width=\linewidth]{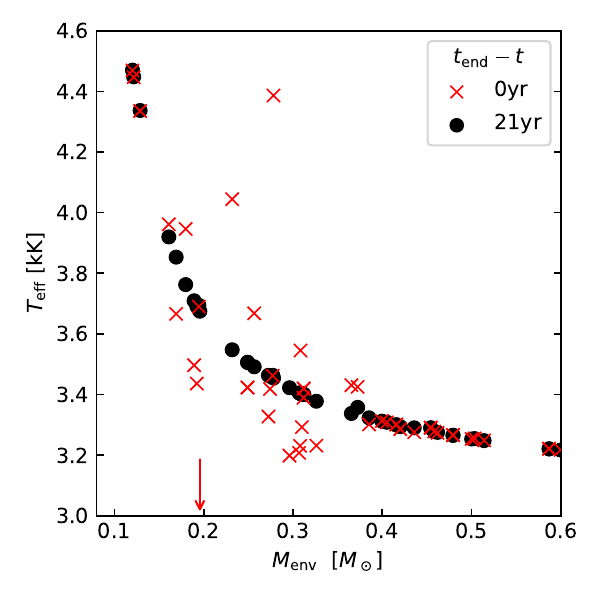}
    \caption{Zoomed-in section of \Fig{fig:Rad_Teff_vs_Menv_preCC}, where in black we report the values of the radius (\EDITL{top}) and effective temperature (\EDITL{bottom}) of the primary star 21 years prior to the end of the run and in red those at the end of the calculation.}
    \label{fig:R_Teff_preCC_smooth}
\end{figure}

After the depletion of carbon in the core, successive burning stages develop off-center flames, which often result in convergence issues. This problem is more pronounced in models with a smaller helium core, where the burning flames develop further off-center. To help convergence and speed up the calculation, some numerical settings (e.g., mesh and timestep refinement) are altered after core-carbon depletion, as well as some mixing settings, namely the thermohaline mixing efficiency is increased to 2 and a small step-overshooting is introduced across the metal-burning regions with   $\alpha_\text{ov}=0.008$. These minor changes have a negligible effect on the overall core and envelope structure at the end.
 
  The termination point for the models at $\qi=0.95$ is core collapse, which is defined in MESA as the moment when the iron core collapses at velocities above $1\,000\kms$. Of these, some exhibit numerical instabilities in the final stages, and the end of the run is found after the end of Si-burning, which is only a few hours ahead of core collapse. This guarantees that the structure of the star outside the Si/Fe-rich core will be practically unchanged by the time of the explosion.
 To capture the collapse of the iron core, the hydrodynamic terms are switched on inside the helium core. Due to numerical issues in some models when they become RSGs, the hydrodynamic terms are also turned on for the inner regions of the envelope. The usage of the hydrodynamic terms throughout the entire envelope was not pursued. Even though this would have allowed proper modeling of radial pulsations, this would have been outside the scope of this work. For all other models with $\qi\leq0.90$, the termination is set to core silicon depletion.

However, some models still fail to reach Si burning or do so while exhibiting some anomalous behavior. We will discuss these numerical issues, depending on whether they occurred on the primary star or the secondary.    

\subsection{Numerical instabilities of the primary star model}\label{sec:App_primary}
 A limited set of models (with $M_\text{env}\sim0.15-0.40\Msun$) exhibited strong oscillations of surface quantities (such as radius, luminosity and effective temperature) with periods of $12-20$ days between the end of core-carbon burning and the onset of neon burning. Given such short characteristic timescales, which are almost an order of magnitude smaller than the dynamical or thermal timescale of the models, we believe that these oscillations arise from numerical instabilities rather than some physical process. This instability would arise at most 20 years before the end of the run (which, in all but one model, is the moment of core collapse), and with such small timescales the effect on mass-loss is negligible. However, if one examines the models before the oscillations would set in, all the models in \Fig{fig:Rad_Teff_vs_Menv_preCC} would align on one trend line, and the noise would disappear (cf. \Fig{fig:R_Teff_preCC_smooth}).

\subsection{Numerical instabilities of the secondary star model}\label{sec:App_secondary}
\EDIT{When the secondary star accretes He-enhanced material while critically rotating during \case C RLOF (which occurs exclusively in the \case B+C systems)}, its outer layers exhibit numerical instabilities that often result in a termination of the run during the final stages of the primary star's evolution. This seems to arise due to the abrupt change in its surface composition which results in a change in the radius of the star while critically rotating. Since our main focus is on the primary star rather than the secondary star, we stop following the detailed evolution of the secondary (i.e., it is set to a point mass and mass transfer is artificially set as inefficient)  if it reaches critical rotation and the primary has reduced the mass fraction of carbon in the center below $10^{-5}$. This does not affect our primary star evolution, as mass transfer is still accounted for and only $\lesssim40$ years are left before the collapse. Specifically, the set of models with $P_\text i =1000,\,\ldots,1995\days$ generally suffer from this. \EDIT{The core evolution of the secondary star is also not expected to be significantly affected by the almost negligible amount of mass that would otherwise be accreted during this time span. }

\section{Discussion of inherent uncertainties}\label{sec:APP_uncertanties}

\subsection{Winds from RSGs}
\label{sec:App_uncertanties_winds}

In many of the binary models investigated above,
the primary star ends its life as a RSG with a highly reduced envelope mass, i.e., for our chosen initial primary mass of $\sim$12.6$\mso$, many models end up with
an envelope mass well below $1\mso$, while a single star would retain an envelope of $\sim 7\mso$.

The driving mechanism of RSG winds is not yet well understood. Therefore,
we cannot predict the effect of the reduced mass on the mass-loss rate. 
In our RSG stellar models, for which we use the empirical mass-loss recipe of 
\cite{Nieuw_wind},
a mass dependence of the mass-loss rate is included, but it predicts weaker 
winds for less massive stars with the same radius and luminosity ($\dot M \sim M^{0.16}$). 

However, since radius, effective temperature and luminosity of the partially stripped models are similar to that of the single star model, one might
argue that the mass-loss rate must be enhanced, due to the reduced gravity and increased Eddington factor, as envisioned in the classical
Reimers-law \citep[$\dot M \sim R L / M$, ][]{Reimers_windRSG}. The Reimers-law is just based on dimensional arguments
and lacks a physical basis. However, the latter is provided by \cite{SC_ReimersLaw2}, who also refine it through additional dependencies
of the mass-loss rate on the effective temperature and on the 
RSG surface gravity, leaving the mass dependence, and found that the result is nearly unaffected.
The implication is that the stellar wind mass-loss rate of partially stripped RSGs may be up to three times higher than those of ordinary RSGs of the same
luminosity and radius, and that the wind mass-loss rates of our partially stripped RSG models (cf., Figs.\,\ref{fig:MdotC_q0p95} and \ref{fig:MdotC_dq_1413_1995} and \Tab{table:data}) may be underestimated by
a similar factor.

In addition to that, stellar models generally increase their radius and luminosity considerably after core helium exhaustion, due to helium shell burning. This feature (which enables \Case C mass transfer) increases the $L/M$-ratio by another factor of $\sim 3$ over the corresponding value during core helium burning in our models.
We conclude that the RSG mass-loss rates of our pre-SN models may be underestimated by up to a factor of 10,
in addition to the inherent uncertainty of the empirical RSG mass-loss rate recipe.

\subsection{Winds of hot stripped stars}\label{sec:App_uncertanties_winds_WR}
Wind mass-loss in hot stripped stars can play a role in the evolution of systems which have undergone \case B mass transfer \citep{Gilkis_Wind}. A lower mass-loss rate in this phase would imply a higher final envelope mass and perhaps prevent full stripping for the tighter systems.  
The work of \cite{Gilkis_fits} shows that for \EDIT{partially }stripped-envelope SN progenitors which have been imaged before the explosion, the best fitting models favor wind mass-loss rates as described in \citet{Vink_WR}, rather than those of \citet{NugisLamers2000}, which are adopted here.
This may imply that \EDIT{for} our models which perform blue-loops during core helium burning (Sect.\,\ref{sec:coreHeburn}), less mass-loss may be expected, which would shift the parameter space for \Type Ib SNe towards initially tighter systems, thereby extending the parameter space for \type IIb progenitors.

\subsection{Mass-transfer efficiency}\label{sec:APP_mass_transfer_efficiency}
Our models are run with a \EDIT{variable} mass transfer efficiency, which is high as long as the accretor is not critically rotating, at which point drops to \EDIT{zero}. \EDIT{This results in a very low mean mass transfer efficiency in every model transferring significant amounts of material,} as \EDIT{the} secondary is spun-up after accreting only a fraction of a solar mass.

In our models, a higher mass transfer efficiency would have a number of effects. Firstly, increasing the efficiency means that less angular momentum is lost from the system during mass transfer, leading to wider orbits. Secondly, the mass ratios would increase, leading to tighter Roche lobes for the primary stars. The overall result of these two competing effects is that the primary star removes more mass compared to the case of inefficient mass transfer \citep{Claeys_b, Ouchi2017_IIb_RSG_progenitors}. \EDIT{Owing to the smaller Roche lobe of the primary, this can also increase the likelihood of mass transfer turning unstable, according to our definition.} This would similarly affect systems undergoing \case B RLOF as well as systems exhibiting only \case C. However, the effect on \case C RLOF after having already undergone \case B is less clear, as the orbit is wider while the star retains more of its envelope.

If both the accretion efficiency and the mass transfer rate are high, then the secondary could find it difficult to thermally relax during accretion (\citealt{Braun_Langer_95}, Schürmann et al., in prep.). This would lead to an expansion and a higher likelihood of filling its Roche lobe, which could lead to the development of a common envelope. 

\subsection{  \EDIT{Loss of mass and} angular momentum \EDIT{from the system}}  \label{sec:APP_angular_momentum_loss}

    Mass that is not accreted onto the secondary is assumed to be \EDIT{ejected} from the binary and it is assumed that it carries \EDIT{away, on average,} the specific orbital angular momentum of the secondary star, as commonly used in other works \citep[e.g., ][]{Claeys_b, Gilkis_Wind, Long_Binary_IIb_2022, Klencki_partialstripping_2022}. However, this is uncertain. A larger angular momentum loss than that used in our models would cause the orbit to widen less during mass transfer, forcing the primary to shed more mass. During \EDIT{the first mass transfer event}, this would result in more \EDIT{stripping} and tighter orbits, making it difficult to determine the behavior of a second RLOF. If the system only exhibited \case C mass transfer, then the higher mass transfer rates would also produce a denser CSM. \EDIT{Additionally, this would likely result in more systems developing unstable mass transfer.} Less angular momentum loss would have the opposite effect.

\EDIT{The work from \citet{Podsiadlowski_massive_star_binary_interaction_1992} summarizes how different combinations of assumptions in the accretion efficiency and angular-momentum losses affect the final envelope of the primary star. Here, they show that for binaries with $\qi=1, 0.75$ and $0.5$ and $M_\mathrm{1,i}=12\Msun$ (cf. their Fig.\,6-8) the final envelope mass is highly sensitive to the assumptions of angular-momentum losses for inefficient mass transfer, which becomes less significant for higher accretion efficiencies. Although a direct comparison with their models is not possible, due to different initialization and physics, 
their results nevertheless indicate that different assumptions on the angular-momentum losses will inevitably result in significant differences in the final envelope masses, possibly shrinking or widening our parameter space. }

\subsection{Stability of mass transfer}\label{sec:APP_RLOFstability}
\subsubsection{\EDIT{The effect of different mass-transfer stability criteria}}
In this paper, the criterion for unstable mass transfer is determined based on the amount of mass lost by the primary star while expanding beyond its outer Lagrangian point. 
This amount was set to $1\Msun$, which is an arbitrary choice. Setting this threshold to a lower (higher) value would result in the boundary between systems undergoing stable and unstable mass transfer shifting towards higher (lower) $\qi$. However, even the extreme case, \EDITL{that is} setting this threshold to zero \EDIT{\citep[which then coincides with the criterion proposed in][ whereby a system would be considered undergoing unstable mass transfer as soon as the primary star overflows the outer Lagrangian point]{Pavlovskii_Ivanova_2015_MT_from_Giants}}, only adds a few more models towards higher $\qi$.

\EDIT{Another marker for the onset of unstable mass transfer might be the so-called dynamical-timescale mass transfer \citep{Ivanova_book}. Here, the plunge-in phase is expected to occur once changes to the orbital configuration or the donor mass have timescales comparable to that of the orbital period. In our models, if we adopt the same threshold parameter as in \citet{Ivanova_book} and \citet{Temmink_2023_stability_of_MT}, then all \case C models with $\qi\leq0.65$ and \case B models with $\qi\leq0.70$ would undergo unstable mass transfer.}   

A different approach was used in \cite{Pablo_Kolb} , where the onset of unstable mass transfer is set when $\dot M_\text{RLOF}>1\msoy$. \EDIT{This criterion would result in more models being labelled as unstable, as all \case C models models with $\qi\leq 0.60$ and most of the wide \case B models with $\qi\leq.80$ reach this mass transfer rate threshold.}

\subsubsection{\EDIT{The effect of different mass-transfer schemes}}

\EDIT{We cannot discuss the stability of mass transfer without also mentioning the effect of different mass transfer schemes, as there exist more recent alternatives to, or improvements of, the scheme from \citet{Kolb_scheme} for the scope of the study of the systems in this work.}

\EDIT{One such example is the recent improvement to the scheme of \citet{Kolb_scheme} introduced in \citet{Pablo_Kolb} that better accounts for mass transfer by including radiation pressure and a better approximation to the Roche potential around the L1 point. This scheme results in a slightly more compact donor during mass transfer and higher mass transfer rates compared to the original scheme. The first effect would result in more of our systems undergoing stable mass transfer according to our criterion and that of \cite[][]{Pavlovskii_Ivanova_2015_MT_from_Giants}, while the latter would do the exact opposite when adopting criteria based on limits on the mass transfer rate \citep[e.g., ][]{Ivanova_book, Temmink_2023_stability_of_MT}. Another important example is the scheme developed in \cite{Cehula_Pejcha_2023_MTscheme}, which offers, under different circumstances, higher or lower mass transfer rates than that of \cite{Kolb_scheme}.}

\EDIT{This highlights that the adoption of different mass transfer schemes in the calculations has important consequences on the expected stability of mass transfer, which can have different outcomes depending on the criterion adopted. }

\subsection{Common-envelope evolution}
The phase of CE evolution following the onset of unstable mass transfer is still poorly understood.
The energy criterion requires knowledge of the efficiency with which the orbital energy is used to unbind the common envelope, which is encapsulated in the variable $\alpha_\text{CE}$ which we set to one. 
Lower values would result in tighter post-CE evolution orbits and more of the CE systems fulfilling the merger criterion as more orbital energy is required to unbind the same amount of envelope.
This does not affect the systems undergoing unstable \case B mass transfer, since in these systems a merger occurs even with $\alpha_\text{CE}=1$, while for the \Case C systems this would shift the boundary between possible common-envelope ejection and possible merger systems further towards higher $\qi$.
At the same time, the value of $\lambda$ is also a source of uncertainty, as it is evaluated at the onset of unstable mass transfer and does not take into account the changes in the density structure and thermodynamic properties of the CE during CE evolution. 

Finally, the time sequence of events during a CE evolution is also subject to many uncertainties. For example, during the plunge-in phase, the CE may not always become completely unbound \citep{Lau_Hydro_12M}, and it may fall back into the binary system during the in-spiral phase \citep{Fallback_CEE}. 
This may result in some hydrogen being reaccreted onto the primary star, which would strongly affect the SN features. 
Furthermore, in some 3D-hydrodynamical simulations, the secondary star is swallowed by the primary's envelope and then accreted onto the primary's core due to tidal disruption \citep[][]{Betelgeuse_merger}, while in others \citep[][]{Lau_Hydro_12M} parts of the CE may be unbound by the time tidal-disruption occurs.  
It is also important to note that the material that is ejected during CE evolution is not leaving the system isotropically but has a preference towards the orbital plane \citep[e.g.,][]{Fallback_CEE, Lau_Hydro_12M, Gagnier_post_plunge_in}.

\section{Definition of envelope mass}\label{sec:App_core_envelope}
\begin{figure*}
   \resizebox{\hsize}{!}{
        \includegraphics[width=\textwidth]{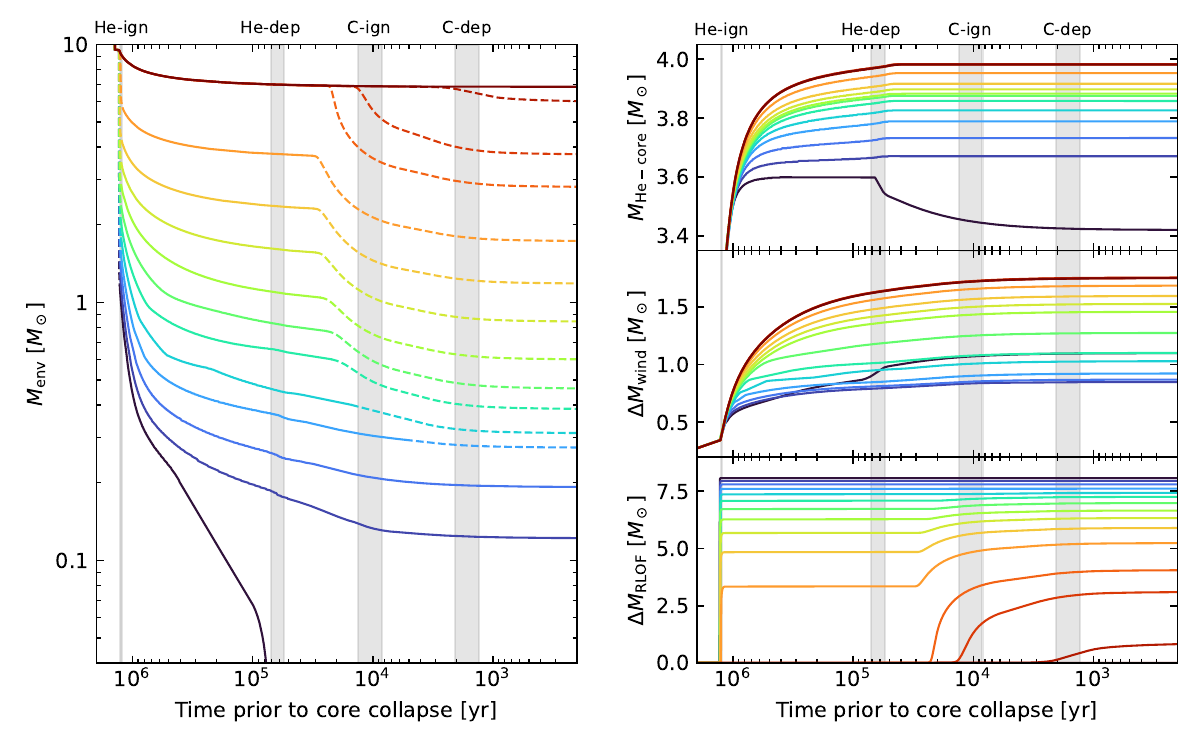}}
      \caption{Evolution of the envelope mass of the primary star in the models with $\qi=0.95$ as a function of time prior to core collapse (left) and the different contributors to its removal (right): growth of the helium core (top), mass-loss by winds (middle) and mass-loss by mass-transfer (bottom). The colors, dashing, and shading are the same as in \Fig{fig:R_vs_t_and_R_vs_Menv}.}
         \label{fig:Menv_vs_t}
\end{figure*}

\begin{figure}
    \includegraphics[width=0.95\columnwidth]{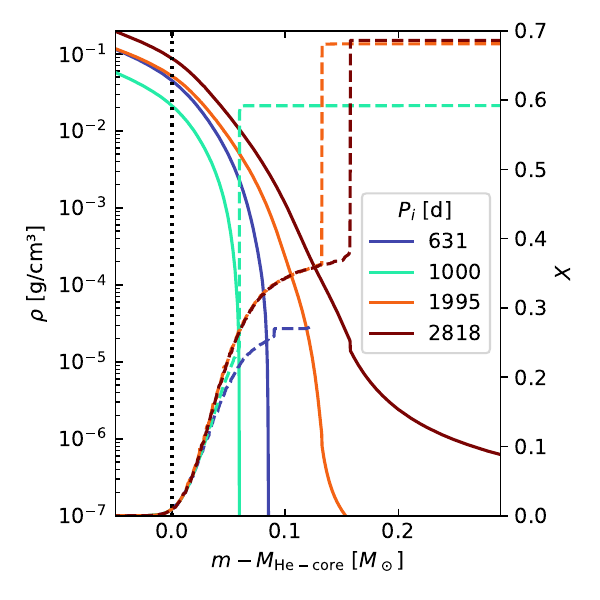}
    \caption{Close-up view of the density (solid) and hydrogen (dashed) profile around the core-envelope boundary in some of the models with $\qi=0.95$ at the time of core collapse. The dotted line indicates the location of the helium core mass as evaluated from MESA. }
    \label{fig:core_env_bound}
\end{figure}
\begin{figure}
    \includegraphics[width=0.95\columnwidth]{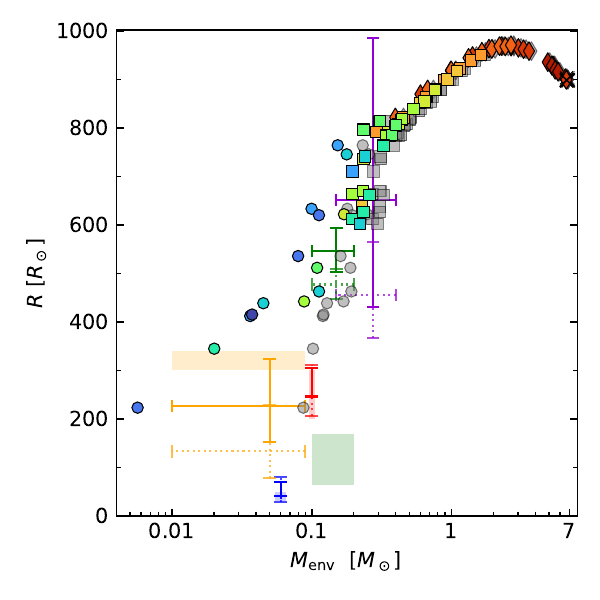}
    \caption{Same as \Fig{fig:Rad_Teff_vs_Menv_preCC} but with the envelope mass estimated by only considering the mass of the star with $\rho\leq10^{-5}\gcc$ (colored markers), instead of that with $X>0.01$ (gray markers).}
    \label{fig:R_vs_Menvext}
\end{figure}

Stellar modeling in MESA refers to helium core mass $M_\text{He-core}$ as the mass coordinate of the star that has negligible amounts of hydrogen, the threshold of which is set in this work as $X=0.01$. Using this definition, the envelope mass will be defined as the remaining mass of the star found outside this mesh, \EDITL{that is} $M-M_\text{He-core}$. The time evolution of the envelope mass is linked to the mass-loss of the star, via winds or mass transfer, as well as the growth of the helium core (cf. \Fig{fig:Menv_vs_t}).

This definition of the helium core has practical uses, as the layers just above this should have enough hydrogen to undergo convective hydrogen burning. This implies that the steep density gradient associated with the extended envelope will be found at points further out than the boundary prescribed by the code (cf. \Fig{fig:core_env_bound}). 

It is important to note that there is a difference between this definition of the envelope mass and the one used in light-curve modeling, which is instead based on thresholds in density.

In light-curve modeling, only the `loosely' bound envelope, \EDITL{that is} the meshes found at large radii, contribute sensitively to the light curve, regardless of their composition.
The region surrounding the core exhibits a relatively steep density gradient, which would jump from $\simgr10^{-2}\gcc$ to as low as $10^{-5}$ in a matter of $0.05-0.15\Msun$ (cf. \Fig{fig:core_env_bound}). Assuming that regions with densities $\rho\simgr10^{-4}$ do not contribute to the signal in the light-curve, as they are found in our models within $6\Rsun$, then it is safe to assume that the extended envelope mass of our models actually is $0.05-0.15\Msun$ less compared to our estimate from the hydrogen-abundance threshold. 

While this difference is negligible for stars with envelope masses greater than $1\Msun$, it can become important for our models with smaller estimated envelope masses (cf \Fig{fig:R_vs_Menvext}). In these models, this difference results in a significant fraction of the estimated envelope mass to actually be found in the dense layers surrounding the core and thus do not contribute to the shape of the light-curve. Finally, the sudden drop in radius at $M_\mathrm{env}\simle 0.3\Msun$ in \Fig{fig:Rad_Teff_vs_Menv_preCC} is now replaced in \Fig{fig:R_vs_Menvext} by a log-linear trend. With this change, our results appear to be more in contrast with the estimates of \SN{2011dh} and especially \SN{2008ax}, while they are still compatible within the errors to the estimates for \SN{2016gkg}.

\section{Comparison with previous works}\label{sec:APP_compare}

Here we compare and contrast the findings of previous studies with the key results of this work, namely the presence and features of partially stripped supergiants, the number of \Type Ib progenitors, \case B+C mass transfer, stable \case C mass transfer, and \Type IIn progenitors.  

\subsection{Partially stripped supergiants}
We find evidence of partially stripped supergiants at solar metallicity, in agreement with the initially wider binary models described \cite{Claeys_b}, \cite{Yoon_IIb_Ib}, \cite{Ouchi2017_IIb_RSG_progenitors}, \cite{Sravan_b, Sravan_b_preexp}, and \cite{Long_Binary_IIb_2022}. These works also generally agree with our finding that there exists some threshold $M_\text H$ which distinguishes RSGs and YSGs and compact progenitors, although the specific thresholds vary between studies. 

We find that a star will swell up to become a supergiant if $M_\text H\simgr0.01\Msun$, and specifically those with $M_\text H\simle 0.02\Msun$ explode as YSGs while those with larger $M_\text H$ do so as RSGs. In \citet{Claeys_b}, which focused on initially wide binaries with high mass ratios and $M_\text{1,i}=15\Msun$, they do not find any compact progenitor, which may be owed to the limited parameter space explored and the wind schemes adopted, while the boundary between YSGs and RSGs is found at $M_\text H\sim0.1\Msun$. In the solar-metallicity binary models from \cite{Sravan_b,Sravan_b_preexp} with different initial masses, and initial orbital parameters, compact progenitors are found for $M_\text H\simle 0.2\Msun$, YSGs for $M_\text H \simle 0.3\Msun$ and RSGs for larger masses. In the models from \citet{Yoon_IIb_Ib} with $M_\text{1,i}=11\Msun$ and $13\Msun$,  they find compact progenitors which retain hydrogen if $M_\text H\simle 0.004$, while the YSG/RSG boundary is found at $M_\text H\sim 0.015\Msun$. These results are still compatible with \citet{Klencki_partialstripping_2022}, where they find that, for models with $M_{1,i}=12\Msun$ and $14\Msun$, even their initially widest systems undergoing \case B mass transfer become fully stripped due to winds during core-helium burning, which can be attributed to the initially lower mass ratio of $\qi=0.60$, which favors much stronger stripping during \case B RLOF.

\subsection{\Type Ib progenitors}\label{sec:App_Ib}
The parameter space for \Type Ib progenitors is sensitive to the adopted wind schemes after \case B interaction \citep{Gilkis_Wind}, and thus there is no clear consensus between different work.

Our work implements the \cite{NugisLamers2000} wind mass-loss in hot and helium-enhanced post-\case B envelopes, and results in full stripping in those that had retained an envelope of $M_\text{env}\simle1\Msun$ following \case B RLOF. This result is qualitatively shared by the works of \cite{Yoon_IIb_Ib}, \cite{Sravan_b}, \cite{Long_Binary_IIb_2022}, and \cite{Klencki_partialstripping_2022} and the NL00 series from \cite{Gilkis_Wind}, which share our wind prescription for hot and helium-enhanced stars. 

A different behavior arises in \citet{Claeys_b}, where they use weaker winds described in \citet{deJager_winds_1988}, and the V17 models from \cite{Gilkis_Wind} which adopted the winds from \cite{Vink_WR}, result in almost no fully stripped post-\case B primaries.

In \cite{Ouchi2017_IIb_RSG_progenitors}, while the adopted winds are similar to ours, the implementations differ, and this results in only their tightest models losing the entire envelope.

\subsection{\Case B+\case C mass transfer}

We find that all models which retained $M_\text{env}\simgr 0.3\Msun$ filled their Roche lobes once again, transferring up to $2\Msun$ of material. This second phase of mass transfer is present in some models in the literature, for example in \cite{1993J_Woosley}, \cite{Claeys_b}, \cite{Yoon_IIb_Ib}, \cite{Long_Binary_IIb_2022}, \cite{Ouchi2017_IIb_RSG_progenitors}, and \cite{Matsuoka_Sawada_BinaryInteraction_IIP_Progenitors}, but there is no systematic investigation of this phase. 

In \cite{Gilkis_Wind}, their parameter space does not cover wide enough binaries to exhibit a second phase of mass transfer like ours, and they instead exhibit a second phase of mass transfer for their initially tighter models run with the wind mass-loss rates from \cite{Vink_WR} for hot and helium enhanced stars. A similar behavior is also found in \cite{Yoon_IIb_Ib} for the models with an artificially reduced wind mass-loss rate. As such, different choices of the wind scheme for post-\case B primaries may add the tightest \case B systems to the pool of systems undergoing a second phase of RLOF. 

In \cite{Ouchi2017_IIb_RSG_progenitors}, evidence for a significant second mass transfer phase is also seen for wider models, regardless of mass-transfer efficiency.

\subsection{\Case C mass transfer}\label{sec:App_comp_caseC}
\case C interaction is an evolutionary phase that has not been systematically investigated in the literature, due to its limited parameter space \citep[see e.g.,][]{Schneider_preSNevo_stripped_2021} and the general idea that fully convective envelopes undergoing mass transfer quickly become unstable, even though it was invoked to explain the progenitor structure for \SN{1993J} \citep{1993J_Podsiadlowski, Maund}. In our work we do find this phase of mass transfer to be stable if $\qi>0.5$. In works like \cite{Claeys_b} and \cite{Sravan_b}, whose purpose was to investigate the parameter space for \Type IIb SNe, there are progenitor models that only undergo \case C mass transfer. However, the stability and the full extent of the parameter space for this phase is not discussed. In \cite{Pablo_Kolb}, where they investigate main-sequence stars of $30\Msun$ with black-hole companions of up to $15\Msun$ with a modified version of the Kolb-scheme, they do find that stable \case C mass transfer is found for $\qi\simgr 0.5$ and for models which filled their Roche lobes close to core collapse (see Sect.\,\ref{sec:outliers}). 

The work from \cite{Ouchi2017_IIb_RSG_progenitors} shows example calculations for mass transfer in wide binaries of slightly larger initial mass to ours ($16\Msun$). Comparing their Fig.\,10 and Fig.\,11 with our  \Fig{fig:MdotC_q0p95} confirms the essentials for our models undergoing stable \Case C mass transfer, while some differences in the detailed time dependence of the mass transfer rate may depend on the different implemented physics and parameters (e.g., their models are non-rotating and adopt $\alpha_\text{MLT}=2$),  as well as the exact initial binary configuration. These differences also result in the boundaries between models undergoing \case B+\case C mass transfer, \case C mass transfer and no mass transfer at all to be found at initially tighter systems compared to our own. Similarly, the recent work of \cite{Matsuoka_Sawada_BinaryInteraction_IIP_Progenitors}, which expands on the work of \cite{Ouchi2017_IIb_RSG_progenitors} to models with a primary initial mass of $12\Msun$, also shows qualitative agreement with our models undergoing \case C mass transfer (cf., their Fig.\,1 and Fig.\,4 with our  \Fig{fig:MdotC_q0p95}).

\end{appendix}

\end{document}